\documentclass[dvipsnames, twocolumn]{aastex63}

\newcommand{\ee}[1]{\mbox{${} \times 10^{#1}$}}
\newcommand{\eten}[1]{\mbox{$10^{#1}$}}
\newcommand{\msun}{\mbox{M$_\odot$}}
\newcommand{\rsun}{\mbox{R$_\odot$}}%
\newcommand{\as}{\mbox{\arcsec}}
\newcommand{\degree}{\mbox{$^{\circ}$}}

\newcommand{\water}{H$_2$O}
\newcommand{\co}{$^{12}$CO}
\newcommand{\coo}{$\mathrm{^{13}CO}$}
\newcommand{\cooo}{C$^{18}$O}
\newcommand{\coooo}{C$^{17}$O}
\newcommand{\cotwo}{CO$_2$}
\newcommand{\hh}{\mbox{{\rm H}$_2$}}

\newcommand{\hcop}{\mbox{\rm HCO}$^+$}

\usepackage{amsmath}

\received{}
\revised{\today}
\accepted{}
\submitjournal{AJ}

\shorttitle{MAPS XX}
\shortauthors{Schwarz et al.}
\graphicspath{{./}{figures/}}

\begin{document}

\title{Molecules with ALMA at Planet-forming Scales. XX. The Massive Disk Around GM Aurigae}

\correspondingauthor{Kamber R. Schwarz}
\email{schwarz@mpia.de}

\author[0000-0002-6429-9457]{Kamber R. Schwarz}
\altaffiliation{NASA Hubble Fellowship Program Sagan Fellow}
\affiliation{Lunar and Planetary Laboratory, University of Arizona, 1629 E. University Blvd, Tucson, AZ 85721, USA}
\affiliation{Max-Planck-Institut für Astronomie, Königstuhl 17, 69117 Heidel-
berg, Germany}

\author[0000-0002-0150-0125]{Jenny K. Calahan} 
\affiliation{Department of Astronomy, University of Michigan, 323 West Hall, 1085 South University Avenue, Ann Arbor, MI 48109, USA}

\author[0000-0002-0661-7517]{Ke Zhang}
\altaffiliation{NASA Hubble Fellow}
\affiliation{Department of Astronomy, University of Wisconsin-Madison, 
475 N Charter St, Madison, WI 53706}
\affiliation{Department of Astronomy, University of Michigan, 323 West Hall, 1085 South University Avenue, Ann Arbor, MI 48109, USA}

\author[0000-0002-2692-7862]{Felipe Alarc\'on} \affiliation{Department of Astronomy, University of Michigan, 323 West Hall, 1085 South University Avenue, Ann Arbor, MI 48109, USA}

\author[0000-0003-3283-6884]{Yuri Aikawa}
\affiliation{Department of Astronomy, Graduate School of Science, The University of Tokyo, Tokyo 113-0033, Japan}

\author[0000-0003-2253-2270]{Sean M. Andrews} \affiliation{Center for Astrophysics \textbar\ Harvard \& Smithsonian, 60 Garden St., Cambridge, MA 02138, USA}

\author[0000-0001-7258-770X]{Jaehan Bae}
\altaffiliation{NASA Hubble Fellowship Program Sagan Fellow}
\affil{Earth and Planets Laboratory, Carnegie Institution for Science, 5241 Broad Branch Road NW, Washington, DC 20015, USA}
\affiliation{Department of Astronomy, University of Florida, Gainesville, FL 32611, USA}

\author[0000-0003-4179-6394]{Edwin A. Bergin} \affiliation{Department of Astronomy, University of Michigan, 323 West Hall, 1085 South University Avenue, Ann Arbor, MI 48109, USA}

\author[0000-0003-2014-2121]{Alice S. Booth} \affiliation{Leiden Observatory, Leiden University, 2300 RA Leiden, the Netherlands}
\affiliation{School of Physics and Astronomy, University of Leeds, Leeds, UK, LS2 9JT}

\author[0000-0003-4001-3589]{Arthur D. Bosman} \affiliation{Department of Astronomy, University of Michigan, 323 West Hall, 1085 South University Avenue, Ann Arbor, MI 48109, USA}

\author[0000-0002-2700-9676]{Gianni Cataldi}
\affil{National Astronomical Observatory of Japan, Osawa 2-21-1, Mitaka, Tokyo 181-8588, Japan}
\affil{Department of Astronomy, Graduate School of Science, The University of Tokyo, Tokyo 113-0033, Japan}

\author[0000-0003-2076-8001]{L. Ilsedore Cleeves} \affiliation{Department of Astronomy, University of Virginia, 530 McCormick Rd, Charlottesville, VA 22904}

\author[0000-0002-1483-8811]{Ian Czekala}
\affiliation{Department of Astronomy and Astrophysics, 525 Davey Laboratory, The Pennsylvania State University, University Park, PA 16802, USA}
\affiliation{Center for Exoplanets and Habitable Worlds, 525 Davey Laboratory, The Pennsylvania State University, University Park, PA 16802, USA}
\affiliation{Center for Astrostatistics, 525 Davey Laboratory, The Pennsylvania State University, University Park, PA 16802, USA}
\affiliation{Institute for Computational \& Data Sciences, The Pennsylvania State University, University Park, PA 16802, USA}
\affiliation{Department of Astronomy, 501 Campbell Hall, University of California, Berkeley, CA 94720-3411, USA}
\altaffiliation{NASA Hubble Fellowship Program Sagan Fellow}

\author[0000-0001-6947-6072]{Jane Huang}
\altaffiliation{NASA Hubble Fellowship Program Sagan Fellow}
\affiliation{Department of Astronomy, University of Michigan, 323 West Hall, 1085 South University Avenue, Ann Arbor, MI 48109, USA}

\author[0000-0003-1008-1142]{John~D.~Ilee} \affil{School of Physics \& Astronomy, University of Leeds, Leeds LS2 9JT, UK}

\author[0000-0003-1413-1776]{Charles J. Law}\affiliation{Center for Astrophysics \textbar\ Harvard \& Smithsonian, 60 Garden St., Cambridge, MA 02138, USA}

\author[0000-0003-1837-3772]{Romane Le Gal}
\affiliation{Center for Astrophysics \textbar\ Harvard \& Smithsonian, 60 Garden St., Cambridge, MA 02138, USA}
\affiliation{IRAP, Université de Toulouse, CNRS, CNES, UT3, Toulouse, France}
\affiliation{IPAG, Universit\'{e} Grenoble Alpes, CNRS, IPAG, F-38000 Grenoble, France}
\affiliation{IRAM, 300 rue de la piscine, F-38406 Saint-Martin d'H\`{e}res, France}

\author[0000-0002-7616-666X]{Yao Liu} 
\affiliation{Purple Mountain Observatory \& Key Laboratory for Radio Astronomy, Chinese Academy of Sciences, Nanjing 210023, China}

\author[0000-0002-7607-719X]{Feng Long}
\affiliation{Center for Astrophysics \textbar\ Harvard \& Smithsonian, 60 Garden St., Cambridge, MA 02138, USA}

\author[0000-0002-8932-1219]{Ryan A. Loomis}\affiliation{National Radio Astronomy Observatory, 520 Edgemont Rd., Charlottesville, VA 22903, USA}

\author[0000-0003-1283-6262]{Enrique Mac\'ias}
\affiliation{Joint ALMA Observatory, Avenida Alonso de Córdova 3107, Vitacura, Santiago, Chile}
\affiliation{European Southern Observatory, Avenida Alonso de Córdova 3107, Vitacura, Santiago, Chile}

\author[0000-0003-1878-327X]{Melissa McClure}
\affiliation{Leiden Observatory, Leiden University, 2300 RA Leiden, the Netherlands}

\author[0000-0002-1637-7393]{Fran\c cois M\'enard}\affiliation{Univ. Grenoble Alpes, CNRS, IPAG, F-38000 Grenoble, France}

\author[0000-0001-8798-1347]{Karin I. \"Oberg} \affiliation{Center for Astrophysics \textbar\ Harvard \& Smithsonian, 60 Garden St., Cambridge, MA 02138, USA}

\author[0000-0003-1534-5186]{Richard Teague} \affiliation{Center for Astrophysics \textbar\ Harvard \& Smithsonian, 60 Garden St., Cambridge, MA 02138, USA}

\author[0000-0001-7591-1907]{Ewine van Dishoeck}
\affiliation{Leiden Observatory, Leiden University, 2300 RA Leiden, the Netherlands}

\author[0000-0001-6078-786X]{Catherine Walsh}\affiliation{School of Physics and Astronomy, University of Leeds, Leeds, UK, LS2 9JT}

\author[0000-0003-1526-7587]{David J. Wilner} \affiliation{Center for Astrophysics \textbar\ Harvard \& Smithsonian, 60 Garden St., Cambridge, MA 02138, USA}


\begin{abstract}
Gas mass remains one of the most difficult protoplanetary disk properties to constrain. With much of the protoplanetary disk too cold for the main gas constituent, \hh, to emit, alternative tracers such as dust, CO, or the \hh\ isotopolog HD are used. However, relying on disk mass measurements from any single tracer requires assumptions about the tracer's abundance relative to \hh\ and the disk temperature structure. 
Using new Atacama Large
Millimeter/submillimeter Array (ALMA) observations from the Molecules with ALMA at Planet-forming Scales (MAPS) ALMA Large Program as well as archival ALMA observations, we construct a disk physical/chemical model of the protoplanetary disk GM Aur. Our model is in good agreement with the spatially resolved CO isotopolog emission from eleven rotational transitions with spatial resolution ranging from $0.15\as$ to 0.46\as (24-73 au at 159 pc) and the spatially unresolved HD $J=1-0$ detection from Herschel. 
Our best-fit model favors a cold protoplanetary disk with a total gas mass of approximately 0.2 \msun, a factor of 10 reduction in CO gas inside roughly 100 au and a factor of 100 reduction outside of 100 au. Despite its large mass, the disk appears to be on the whole gravitationally stable based on the derived Toomre $Q$ parameter. However, the region between 70 and 100 au, corresponding to one of the millimeter dust rings, is close to being unstable based on the calculated Toomre $Q$ of $<1.7$. This paper is part of the MAPS special issue of the Astrophysical Journal Supplement.

\end{abstract}


\section{Introduction} \label{sec:intro}
Protoplanetary disk mass is a fundamental property influencing virtually all aspects of the disk's evolution and the resulting planetary system. 
It sets a limit on the mass available to forming planets and determines the mechanisms that shape the final system architecture.
Gravitational instability (GI) in a protoplanetary disk can result in the formation of massive companions at separations of hundreds of astronomical units from the central star \citep{Boss97}.
Gravitational collapse at early stages results in the formation of close multiple-star systems \citep{Tobin18}. In other cases, GI can result in the formation of massive planets at wide separation, such as the giant planets proposed to exist in the HD 163296 disk \citep{Isella16,Liu18,Pinte18,Teague18}. 

In a recent study of the HD 163296 system, \citet{Booth19} found a total disk mass of 0.31 \msun\ based on observations of the optically thin CO isotopolog $^{13}$C$^{17}$O. They further found that the disk currently is stable against gravitational collapse, though they noted the disk may have been more massive, and thus unstable, in the past, having had ample time to accrete mass onto the central star. Analysis of the HL Tau disk using the same $^{13}$C$^{17}$O transition indicates this much younger system surrounded by a residual envelope has a lower total disk mass of 0.2 \msun\ but is likely unstable at radii from 50-110 au \citep{Booth20}, spanning several of the rings and dark bands observed in the millimeter continuum \citep{ALMA15}.

Disk gas mass remains one of the most difficult parameters to constrain. This is because the dominant gas component, \hh, does not readily emit throughout most of the disk due to its lack of a permanent dipole moment. Instead, trace species such as CO and dust are used to extrapolate to a total gas mass. 
However, each tracer relies on assumptions that may not be applicable to protoplanetary disks (see \citet{Bergin17} for a review). 

Converting dust continuum emission to a total dust mass requires assumptions about the grain size distribution, dust composition, scattering, and dust temperature. The dust mass is often then converted to a gas mass assuming a gas-to-dust mass ratio of 100, as measured for the interstellar medium (ISM). However, several processes can change the derived ratio, including differential radial drift for grains of different sizes, dust growth beyond the observable range, accretion onto the central star, and photoevaporative winds. Additionally, observations show that the gas disk often extends far beyond the millimeter dust grains \citep{Guilloteau98,Najita18,Facchini19,Trapman19}.

CO column densities derived from optically thin emission can be converted into a total gas mass assuming a CO/\hh\ ratio of $\sim \eten{-4}$ based on ISM values and correcting for the effects of CO freeze-out in the cold disk midplane and CO photodissociation in the surface layers \citep{Miotello14,Williams14}.
However, surveys of protoplanetary disks consistently find a discrepancy between dust-derived and CO-derived disk gas masses, with the CO-based measurements generally indicating a lower mass \citep{Ansdell16,Long17}.
One potential explanation is that gas-phase CO abundance has been further reduced by processes beyond freeze-out and photodissociation, resulting in an underestimation of the total gas. These processes could be physical, such as vertical mixing, which preferentially traps CO in the cold disk midplane \citep{Xu17,Krijt18}, or chemical, with CO processed into other, less emissive species in either the gas or ice \citep{Yu16,Bosman18,Eistrup18,Schwarz18}. 

Alternative tracers are needed to determine the true gas mass in protoplanetary disks. One approach is to use the outer radius of dust emission at different wavelengths to constrain the rate of radial drift \citep{Powell17}. Masses derived using this technique are significantly larger than those derived from dust or CO line emission \citep{Powell19}. However, as this analysis does not yet consider disk substructure, which would impact drift timescales, these values are likely upper limits to the true gas mass.

Another approach is to derive the \hh\ mass from observations of the \hh\ isotopolog HD. The HD/\hh\ ratio of 3\ee{-5} is not subject to the same processes that can change the CO/\hh\ and gas-to-dust mass ratios \citep{Linsky98}. The HD $J=1-0$ transition was detected in three protoplanetary disks using the {\it Herschel Space Observatory}, including MAPS target GM Aur \citep{Bergin13,McClure16}. Upper limits exist for an additional nineteen systems \citep{McClure16,Kama20}. 

Initial analysis of the HD detections has yielded large disk gas masses, more in line with those derived from dust than with those derived from CO \citep{McClure16,Trapman17}. However, the range of disk masses consistent with the observed HD emission strength can be quite large, in some cases spanning more than an order of magnitude.
This is due to the strong degeneracy between HD abundance and gas temperature in contributing to the observed HD emission strength. Further, due to the high $J=1$ upper state energy, the ground state transition of HD does not emit at temperatures lower than roughly 20~K \citep{Bergin13}. Knowledge of the gas temperature structure in the disk from, e.g., spatially resolved CO observations, can reduce the uncertainty on the HD derived gas mass from over an order of magnitude to approximately a factor of two \citep{Trapman17}. 

Subsequent analysis of the HD toward one source, TW Hya, has used observations of CO isotopologs to constrain the gas temperature and, when combined with HD, the gas mass \citep{Favre13,Schwarz16,Zhang17}. Key results include the fact that HD emits primarily within the inner 20 au of the TW Hya disk, with a gas-to-dust mass ratio of 140 in this region, and a CO/\hh\ abundance ranging from $<\eten{-6}$ in the outer disk to greater than \eten{-5} inside the CO snowline, indicating an overall depletion of gas-phase CO in TW Hya as compared to ISM values. Additionally, \citet{Calahan21} demonstrated a wide range of CO abundance and total gas mass are able to reproduce the observed CO emission profiles, while the additional constraint of the HD line provides a way to break this degeneracy.

In this paper we focus on the disk around GM Aur. GM Aur is a 1.1 \msun\ star hosting a well-known transition disk at a distance of 159 pc \citep{Calvet05,Hughes09,Gaia18,Macias18}. We use the CO isotoplogue observations of GM Aur from the Atacama Large
Millimeter/submillimeter Array (ALMA) Large Program
MAPS, along with the HD detection from Herschel, archival ALMA CO observations, and data on the spectral energy distribution (SED) to construct a 2D thermochemical model of the gas density, temperature, and CO abundance in the GM Aur disk. 
The observations and data reduction process are summarized in Section~\ref{sec:obs}. Section~\ref{sec:model} describes the modeling framework used to fit the data. In Section~\ref{sec:results} we present the results of our modeling study. In Section~\ref{sec:discuss} we discuss what our results reveal about the gas mass, temperature, and CO abundance in the GM Aur disk. Finally, our conclusions are summarized in Section~\ref{sec:summary}.

\section{Observations and Data Reduction} \label{sec:obs}
This study uses the CO isotopolog emission toward GM Aur as part of the MAPS Large Program, covering the \coo, \cooo, and \coooo\ $J$ =  1-0 transitions in Band 3 and the \co, \coo, and \cooo\ $J$ =  2-1 transitions in Band 6. The full details of the calibration and imaging processes are described in \citet{ObergMAPS} and \citet{CzekalaMAPS} respectively.
Additionally, we augment these data with CO isotopologs in Band 7 and Band 9 from the ALMA Cycle 4 program 2016.1.00565.S (PI K. Schwarz), targeting the \coo\ and \cooo\ $J$ =  3-2 and \coo, \cooo\ and \coooo\ $J$ = 6-5 transitions. Observations were obtained in Band 7 on 11 November 2016 with 42 antennas. Observations were obtained in Band 9 on 18 August 2018 with 48 antennas. The continuum data associated with the Band 7 observations was previously analyzed by \citet{Macias18}.

\begin{figure*}
    \centering
    \includegraphics[width = 2\columnwidth]{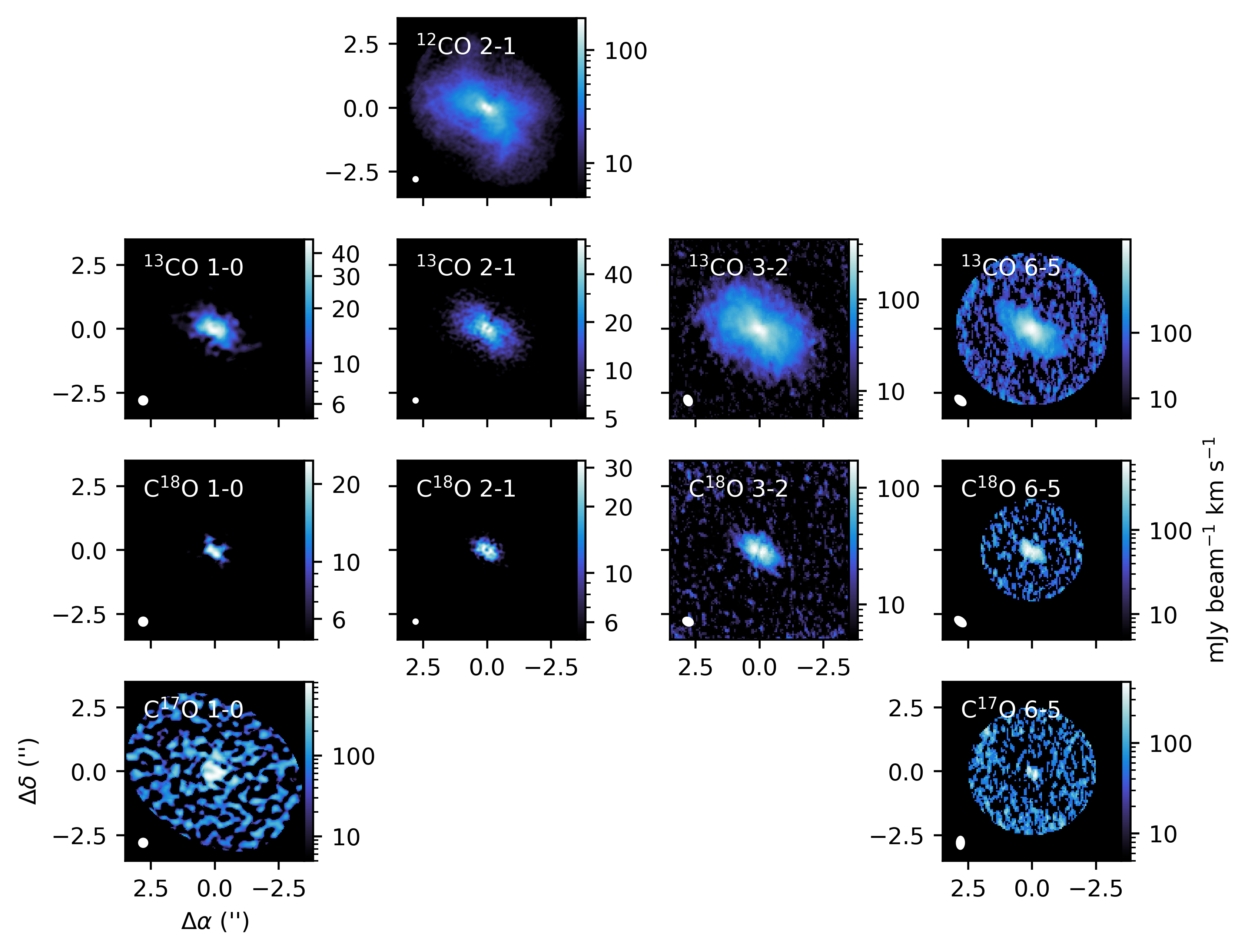}
    \caption{Integrated intensity maps with a logarithmic color stretch for the 11 CO isotopolog rotational transitions considered in this work. The \coooo\ $J=1-0$ map is a sum of the three hyperfine components. The $J=1-0$ and $J=2-1$ transitions were observed as part of MAPS. The $J=3-2$ and 6-5 transitions were observed as part of ALMA program 2016.1.00565.S.}
    \label{mom0}
\end{figure*}

Initial calibration of the archival Band 7 and Band 9 data was carried out by ALMA/NAASC staff using standard procedures.
Additionally, phase and amplitude self-calibration were performed in each band using the continuum visibilities. The \texttt{fixvis} task was used to correct the phase center of each data set. Imaging was performed in \texttt{CASA 5.4} using the imaging scripts developed by \citet{CzekalaMAPS}. The channel widths used in the imaging are 0.2 km s$^{-1}$ in Band 7 and 0.3 km s$^{-1}$ in Band 9. The properties of the final \texttt{CLEAN} images, with a robust weighting of 0.5, are given in Table~\ref{obsprops}, and the moment 0 maps are shown in Figure~\ref{mom0}.  All lines are detected, including the \coooo\ $J$ =  6-5 transition, the first time this transition has been detected toward a protoplanetary disk. 
To help in constraining the disk gas mass we also consider the spatially unresolved HD $J=1-0$ detection from Herschel. \citet{McClure16} reported a $5\sigma$ detection toward GM Aur with a total integrated flux of $2.5\pm0.5 \times 10^{-18}$ W m$^{-2}$.

\begin{deluxetable*}{lcccc}
\tablecaption{GM Aur image parameters}
\label{obsprops}
\tablehead{
\colhead{Molecular Transition} & \colhead{Beam} & \colhead{rms} & \colhead{Channel spacing} \\  
 & \colhead{($\arcsec\times\arcsec$, $\degree$)} & \colhead{(mJy beam$^{-1}$)} & \colhead{(km s$^{-1}$)} & Program ID
}
\startdata 
C$^{18}$O J=1-0 &($0.30\times0.30, 75.0$) & 0.514 & 0.5 & 2018.1.01055.L\\ 
${}^{13}$CO J=1-0 &($0.30\times0.30, 81.1$) & 0.498 & 0.5 & 2018.1.01055.L\\ 
C$^{17}$O J=$1-0$ F=$\frac{3}{2}-\frac{5}{2}$,  &($0.30\times0.30, -29.8$) & 0.629 & 0.5 & 2018.1.01055.L\\ 
C$^{17}$O J=$1-0$ F=$\frac{7}{2}-\frac{5}{2}$,  &($0.30\times0.30, -29.8$) & 0.629 & 0.5 & 2018.1.01055.L\\ 
C$^{17}$O J=$1-0$ F=$\frac{5}{2}-\frac{5}{2}$,  &($0.30\times0.30, -29.8$) & 0.629 & 0.5 & 2018.1.01055.L\\ 
C$^{18}$O J=2-1&($0.15\times0.15, 54.8$) & 0.484 & 0.2 & 2018.1.01055.L\\ 
$^{13}$CO J=2-1&($0.15\times0.15, 72.4$) & 0.660 & 0.2 & 2018.1.01055.L\\ 
CO J=2-1&($0.15\times0.15, 66.3$) & 0.730 & 0.2 & 2018.1.01055.L\\ 
C$^{18}$O J=3-2&($0.37\times0.27, 6.7$) & 9.2 & 0.2 & 2016.1.00565.S\\ 
$^{13}$CO J=3-2&($0.38\times0.27, 7.5$) & 8.3 & 0.2 & 2016.1.00565.S\\ 
C$^{18}$O J=6-5 &($0.46\times0.26, 3.8$) & 48.7 & 0.3 & 2016.1.00565.S\\ 
${}^{13}$CO J=6-5 &($0.46\times0.26, 3.7$) & 56.1 & 0.3 & 2016.1.00565.S\\ 
C$^{17}$O J=6-5 &($0.45\times0.26, 4.7$) & 57.1 & 0.3 & 2016.1.00565.S\\ 
\hline 
\enddata 
\end{deluxetable*}

\section{Methods} \label{sec:model}
\subsection{Physical Model}
Our physical disk model is based on an axisymmetric viscously evolving disk \citep{LyndenBell74,Andrews11}. The surface densities of both gas and dust are described by
\begin{equation}
\Sigma(R)=\Sigma_{c}\left(\frac{R}{R_{c}}\right)^{-\gamma} \exp{\left[-\left(\frac{R}{R_{c}}\right)^{2-\gamma}\right]},
\end{equation}
where $\Sigma_{c}/e$ is the surface density at a characteristic radius $R_{c}$.
The 2D density distribution is then
\begin{equation}
\rho(R,Z) = \frac{\Sigma}{\sqrt{2\pi}R h} \exp{\left[-\frac{1}{2}\left(\frac{Z}{h}\right)^{2}\right]}.
\end{equation}
The scale height $h$ varies as a function of radius such that
\begin{equation}
h(R)=h_{\mathrm{ref}} \left(\frac{R}{R_{\mathrm{ref}}}\right)^{\psi}
\end{equation}
where $h_{\mathrm{ref}}$ is the characteristic scale height at a radius $R_{\mathrm{ref}}$ and $\psi$ is the power-law index characterizing disk flaring.
Our disk includes three populations of matter: gas, small grains, and large grains. The small grains are well mixed with the gas, i.e, have the same scale height, and follow the same surface density profile. The scale height for the large grains is smaller than that of the gas and small grains to mimic vertical settling. Further, while the gas and small grains vary smoothly, the large grain surface density includes several depleted regions corresponding to the gaps seen in continuum emission \citep{Huang20a}. For our large grain model we start with the surface density profile from the model of \citet{Macias18}. This model includes rings and gaps and was able to reproduce both the millimeter and centimeter continuum emission. The initial large grain distribution is then adjusted as described by \citet{ZhangMAPS}.

\subsection{SED Fitting}
Our initial values for the disk model were determined by fitting the SED. The general procedure for modeling the SEDs of all MAPS sources is described in detail by \citet{ZhangMAPS} and we use the same dust surface density models for GM Aur as in that work. Briefly, data is fit using \texttt{RADMC-3D} \citep{Dullemond12}. 
The large grains range in size from 0.005 \micron\ to 1 mm with an MRN distribution $n(a)\propto a^{-3.5}$ \citep{Mathis77} and use the standard dust opacities from \citet{Birnstiel18}, who assumed a dust composed of 20\% \water\ ice, 32.9\% astronomical silicates, 7.4\% troilite, and 39.7\% refractory organics by mass. This differs from 38\% graphite and 62\% silicate composition used in previous SED modeling of GM Aur \citep{Espaillat11}, necessitating some modification of the dust surface density profile from \citet{Macias18}. 
The small grains have a size range of 0.005 \micron\ to 1 \micron\ with an MRN distribution and are assumed to be composed of equal parts silicates and refractory organics by mass. At large scale heights, where most of the dust mass is in small grains, models suggest water is removed from grains via photodesorption \citep{Hogerheijde11}. For a given large and small grain distribution, we used \texttt{RADMC-3D} to compare our model to the observed SED and ALMA continuum image.


\subsection{Thermochemical models}
After using the SED to constrain the dust distribution we pass our disk density model to the 2D thermochemical modeling code \texttt{RAC2D} to model the molecular line emission. \texttt{RAC2D} self-consistently computes the chemistry as well as the evolving balance between heating and cooling in the disk. The modeling framework is described in detail by \citet{Du14}.
For each model run \texttt{RAC2D} first calculates the dust thermal structure, cosmic-ray attenuation, and the radiation field, taking into account photon scatter and absorption. We assume a cosmic-ray ionization rate at the disk surface of 1.38\ee{-18} s$^{-1}$ per \hh, consistent with cosmic-ray modulation by stellar winds \citep{Cleeves13}. Chemical evolution and gas temperature structure are then solved simultaneously. 

\begin{deluxetable}{lc}\label{initialabun}
    \tablewidth{0pt}
    \tablecolumns{2}
    \tablecaption{Standard Initial Chemical Abundances}
    \tablehead{
    &\colhead{Abundance Relative}\\
    &\colhead{to Total H}}
     \startdata
      \hh & \( 5 \times 10^{-1}\) \\
      He & 0.09\\
      CO & \(1.4 \times 10^{-4}\)\\
      N & \(7.5 \times 10^{-5}\)\\
      \water\ ice & \(1.8 \times 10^{-4}\)\\
      S & \(8 \times 10^{-8}\)\\
      Si$^+$ & \(8 \times 10^{-9}\)\\
      Na$^+$ & \(2 \times 10^{-8}\)\\
      Mg$^+$ & \(7 \times 10^{-9}\)\\
      Fe$^+$ & \(3 \times 10^{-9}\)\\
      P & \(3 \times 10^{-9}\)\\
      F & \(2 \times 10^{-8}\)\\
      Cl & \(4 \times 10^{-9}\)\\
     \enddata
\end{deluxetable}

The chemical network is based on the gas-phase network from the UMIST 2006 database \citep{Woodall07} and the grain surface network of \citet{Hasegawa92}. We consider a total of 5830 reactions among 484 species. The chemical network includes two body gas-phase reactions, photodissociation, including Ly-$\alpha$ dissociation of \water\ and OH, adsorption of species onto grain surfaces, thermal desorption, UV photo-desorption, and cosmic-ray induced desorption, as well as a limited network of two body grain surface reactions. 
The default initial chemical composition is given in Table~\ref{initialabun}. 

The chemistry is run for 1 Myr. The main consequence of changing the run time is the amount of chemical processing of CO that takes place, as the gas and dust temperatures converge on much shorter timescales. However, we adjust our initial CO abundance in order to match the observed CO line emission profiles, as described in Section~\ref{sec:parameterstudy}. Thus, the decision on how long to run the chemistry does not impact our final results.

Line radiative transfer calculations assuming local thermal equilibrium are also carried out using \texttt{RAC2D}. 
Line and continuum emission are modeled together using the source properties in Table~\ref{starprops}. The resulting image cubes are then continuum subtracted and convolved with an elliptical Gaussian beam with the same size and orientation as the corresponding observation before comparison to the observations. 
By treating the dust continuum and line emission simultaneously, we account for any extinction of the line emission by the dust.
Our chemical network is not fractionated to include species such as $^{13}$C, $^{18}$O, and deuterium. Instead, isotopolog emission profiles are generated assuming $^{13}$C/$^{12}$C = 69, $^{16}$O/$^{18}$O = 557, $^{18}$O/$^{17}$O = 3.6, and D/H = $1.5\ee{-5}$ as measured in the local ISM \citep{Linsky98,Wilson99}.

\begin{deluxetable}{lcr}\label{starprops}
    \tablewidth{0pt}
    \tablecolumns{2}
    \tablecaption{Source Properties}
    \tablehead{ & Value & Reference}
     \startdata
    distance (pc) & 159 & \citet{Gaia18} \\
    $i$ (\degree) & 53.2 & \citet{Huang20a}\\
    PA (\degree) & 57.2 & \citet{Huang20a}\\
    T$_{\mathrm{eff}}$ (K) & 4350 & \citet{Espaillat11} \\
    M$_*$ (\msun) & 1.1 & \citet{Macias18} \\
    R$_*$ (\rsun) & 1.9 & \citet{Macias18} \\
     \enddata
\end{deluxetable}

\begin{deluxetable}{lccc}\label{paramrange}
    \tablewidth{0pt}
    \tablecolumns{2}
    \tablecaption{Range of parameter values considered}
    \tablehead{ & Gas & Small Dust & Large Dust}
     \startdata
     Mass (\msun) & 0.02-0.41 & $1.03\ee{-4}$ & $5.94\ee{-4}$ \\
     $\mathrm{\psi^{a}}$ & 1-2 & 1-2 & 1-2 \\
    $\mathrm{\gamma}$ & 0.3-1.5 & 0.3-1.5 & \nodata\ \\
     $\mathrm{R_c}$ (au) & 100-176 & 100-176 & \nodata \\
     $\mathrm{h_{ref}}$ (au) & 5-12 & 5-12 & 0.75-12\\     
     $\mathrm{R_{in}}$ (au) & 0.5-27 & \nodata & \nodata \\
     CO/H & $7\ee{-7}-1.4\ee{-3}$ & \nodata & \nodata \\
     \enddata
     \tablenotetext{a}{$\psi$ for the gas and small grains are varied together, while $\psi$ for the large grains is changed independently.}
\end{deluxetable}

\subsection{Parameter study}\label{sec:parameterstudy}
The mass in dust is constrained by the continuum imaging and the SED. In attempting to fit the line emission we limit our parameter study to the total gas mass and the variables that determine the gas and dust density distribution: $\gamma$, $R_{\mathrm{c}}$, $h_{\mathrm{ref}}$, and $\psi$, as well as the initial CO abundance and the gas temperature in the inner disk, as discussed in Section~\ref{sec:results}. In total we generate 145 unique models. The range of parameters considered are given in Table~\ref{paramrange}. Due to the long run times required for each model, it is unrealistic to use a systematic parameter study using, e.g., a Markov Chain Monte Carlo method to find the best-fit model. Instead, parameters are changed one at a time in order to achieve a reasonable fit.

We modify the CO abundance before modeling the chemistry, in contrast with the CO abundance study of \citet{ZhangMAPS}, who modify the CO abundance after modeling the chemistry. Because \texttt{RAC2D} includes line processes when calculating the gas temperature, our choice to remove CO initially impacts the gas temperature structure and, by extension, the strength of the HD $J=1-0$ emission. 
In constructing our initial model we use the same disk parameters as the best-fit model from \citet{ZhangMAPS}.
Initially, we use the CO depletion profile of \citet{ZhangMAPS}. We then run additional thermochemical models with the depletion profile multiplied by a constant factor ranging from 0.1 to 2. 
To quantify how well a given model fits the data we calculate the reduced $\chi^{2}$ for the CO isotopolog emission radial profiles, comparing the model and observed emission at half beam spacing.
We construct a new initial CO abundance profile taking the best fit based on the reduced $\chi^{2}$ at each radius. This model, using an updated CO abundance profile but otherwise using the same parameters as the best-fit model from \citet{ZhangMAPS}, serves as our initial model.

To constrain the model in the vertical direction we compare the extracted emission surfaces from the observations and models for the \cooo\ $J=2-1$, \coo\ $J=2-1$, \co\ $J=2-1$, and \coo\ $J=3-2$ lines. The signal-to-noise ratios of the other transitions are too low to meaningfully constrain the emission height. Emission surfaces for both the observations and models are extracted with the Python package \texttt{disksurf}\footnote{\url{https://github.com/richteague/disksurf}}, using the method presented by \citet{Pinte18}. In regions where the line flux is weak or originating from a large vertical range, i.e., is optically thin, there is greater uncertainty in the derived emission surface for both the models and the observations. For a detailed discussion of this technique as it applies to the MAPS data see \citet{LawMAPS_surf}.
Finally, we compare the results of each model to the total observed HD flux.

\begin{deluxetable}{lccc}\label{initialparams}
    \tablewidth{0pt}
    \tablecolumns{2}
    \tablecaption{Gas and Dust Population Parameters: Initial Model Values}
    \tablehead{ & Gas & Small Dust & Large Dust}
     \startdata
     Mass (\msun) & 0.2 & \( 1.03\ee{-4}\) & \(5.94\ee{-4} \) \\
     \(\psi\) & 1.35 & 1.35  & 1.35\\
     \(\gamma\) & 1.0  & 1.0  & 1.0 \\
     \({\rm R}_{c}\) (au) & 176 &111 &  \nodata \\
     \({\rm R}_{ref}\) (au) & 100 & 100 &  100 \\
     \({\rm h}_{ref}\) (au) & 7.5 & 7.5 &  3.75 \\     
     \({\rm R}_{in}\) (au) & 1.0 & 1.0 & 34 \\
     \({\rm R}_{out}\) (au) & 650 & 650 & 310  \\
     \enddata
\end{deluxetable}

\section{Results}\label{sec:results}
\begin{figure*}[htbp]
    \centering
    \includegraphics[width = 2\columnwidth]{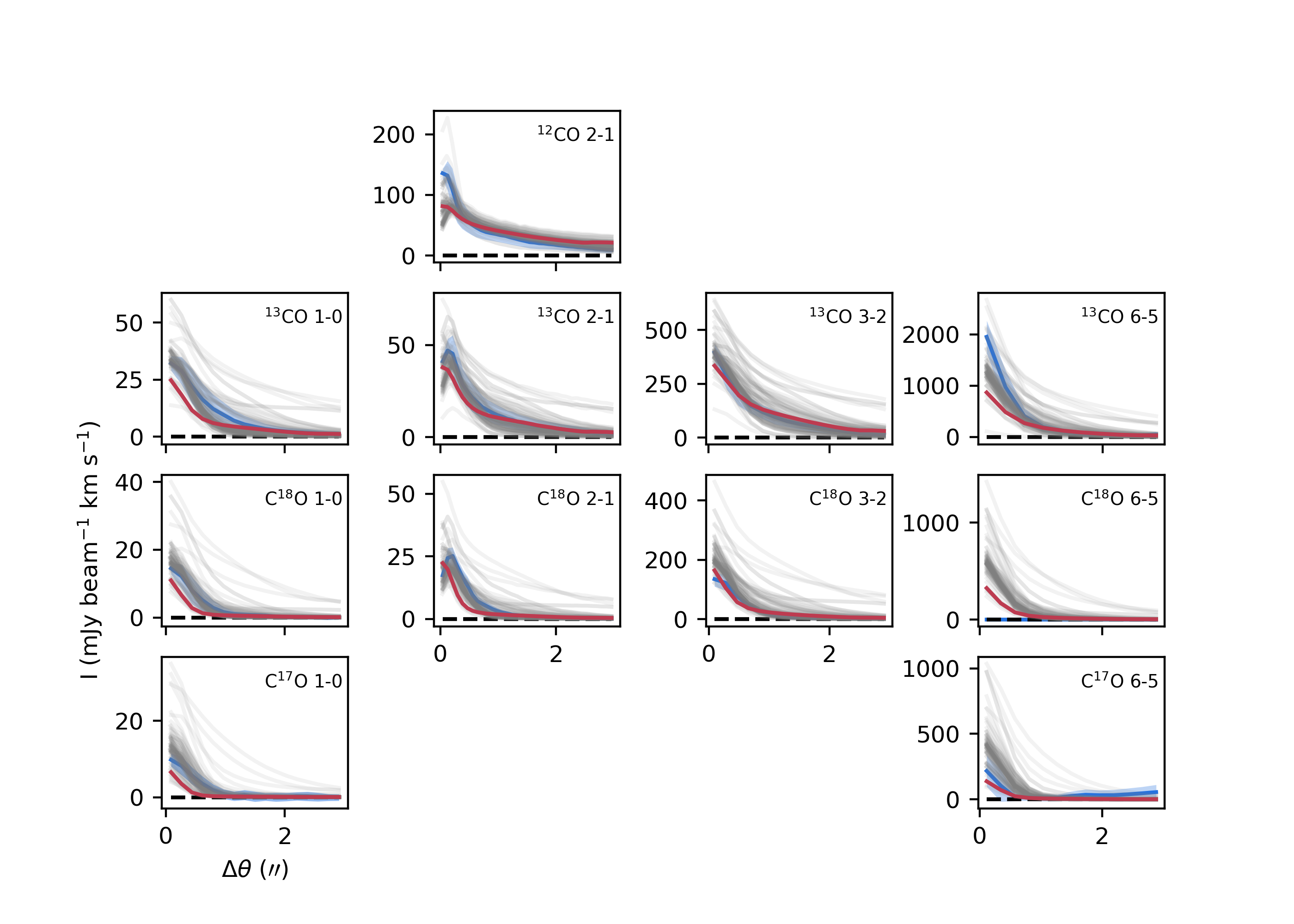}
    \caption{Red lines show the deprojected, azimuthally averaged radial emission profiles for our model using the same disk density parameters as \citet{ZhangMAPS}. Blue lines show the observations. Blue shading indicates the $1\sigma$ uncertainty. Light gray lines show the profiles for all models.}
    \label{profile44}
\end{figure*}

\begin{figure*}[htbp]
    \centering
    \includegraphics[width=2\columnwidth]{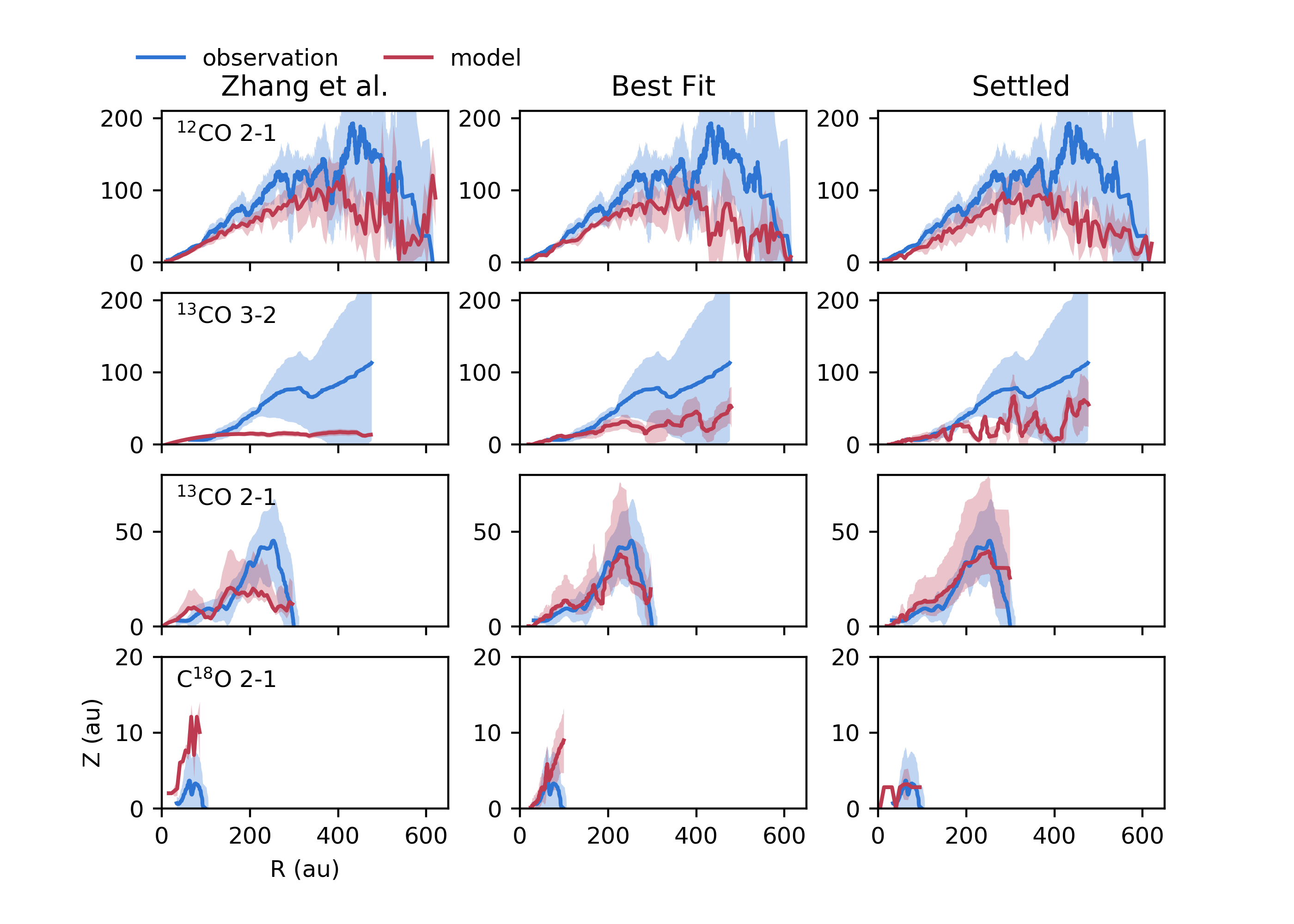}
    \caption{Extracted emission surface profiles for the \co\ 2-1, \coo\ 3-2, \coo\ 2-1, and \cooo\ 2-1 lines in our observations (blue) and three models (red). Shading indicates the uncertainty based on the scatter of points in the extraction before averaging. From left to right the columns show our results for the model using the disk parameters from \citet{ZhangMAPS}, our best-fit model, and our best-fit model with small grain setteling.}
    \label{surfacecombo}
\end{figure*}

The parameters for our initial model, based on the best-fit model of \citet{ZhangMAPS}, are given in Tabel~\ref{initialparams}.
This model fits the radial intensity profiles for the majority of the observed lines within $1\sigma$ outside of 160 au (1\as) (Figure~\ref{profile44}). Inside of 160 au, the model under-predicts the line flux for nearly all transitions. 
The integrated HD $J=1-0$ flux in our initial model is $1.9\ee{-18}$ W m$^{-2}$ compared to the observed $2.5\pm0.5\ee{-18}$ W m$^{-2}$, just below the $1\sigma$ uncertainty.
In the outer disk the model emission surfaces for the \coo\ lines are below the $1\sigma$ uncertainty of the surfaces derived from observations, while for \cooo\ the model over-predicts the emission surface (Figure~\ref{surfacecombo}). 

To raise outer disk emission surfaces in our model, we modify the surface density profile of the gas and small grains by changing $R_c$ and $\gamma$, thus shifting more mass to larger radii, and the disk flaring by adjusting $\psi$. We also consider models with varying total gas mass (see Table~\ref{paramrange}) but find that holding the gas mass at 0.2 \msun\ provides the best fit to the data. 
After adjusting the gas surface density, we modify the initial CO depletion profile as described in Section~\ref{sec:parameterstudy} in order to match the observed radial emission profiles. 
The resulting model brings the \cooo\ 2-1, \coo\ 2-1, and \coo\ 3-2 model emission surfaces within the 1$\sigma$ uncertainty of the surfaces derived from observations throughout much of the disk (Figure~\ref{surfacecombo}). 
Further, the model HD flux increases from 1.9\ee{-18} in our initial model to 2.5\ee{-18} W m$^{-2}$, in excellent agreement with the observed flux of $2.5\pm0.5\ee{-18}$ W m$^{-2}$.  

\begin{figure*}
    \centering
    \includegraphics[width = 2\columnwidth]{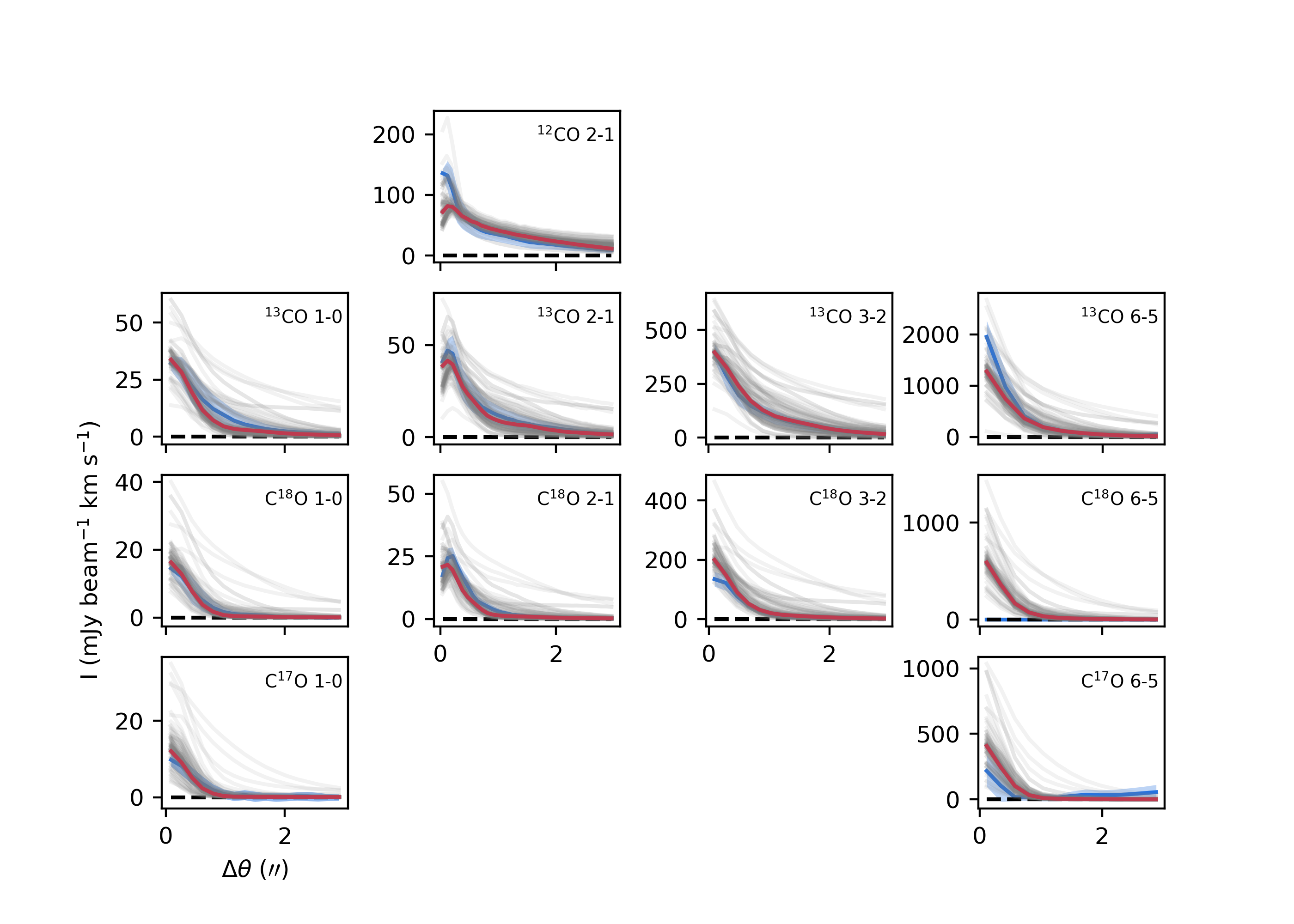}
    \caption{Red lines show the deprojected, azimuthally averaged radial emission profiles for our best-fit model \text{prior to} adjusting the inner disk temperature are. Blue lines show the observations. Blue shading indicates the $1\sigma$ uncertainty. Light gray lines show the profiles for all models.}
    \label{profile62}
\end{figure*}

\begin{figure*}
    \centering
    \includegraphics[width = 2\columnwidth]{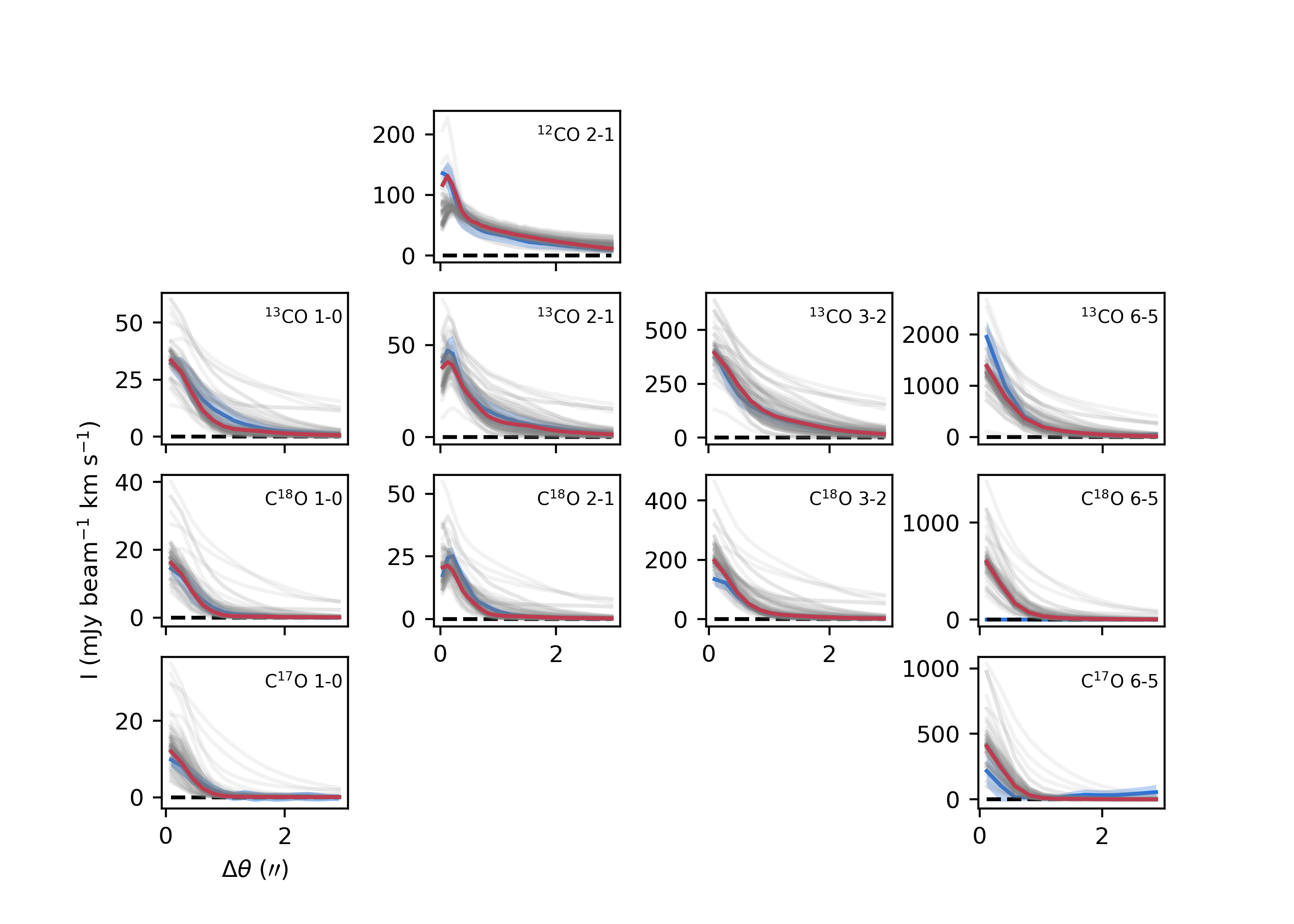}
    \caption{Red lines show the deprojected, azimuthally averaged radial emission profiles for our best-fit model after adjusting the inner disk temperature. Blue lines show the observations. Blue shading indicates the $1\sigma$ uncertainty. Light gray lines show the profiles for all models.}
    \label{profiles}
\end{figure*}
\begin{figure*}
    \centering
    \includegraphics[width = 2\columnwidth]{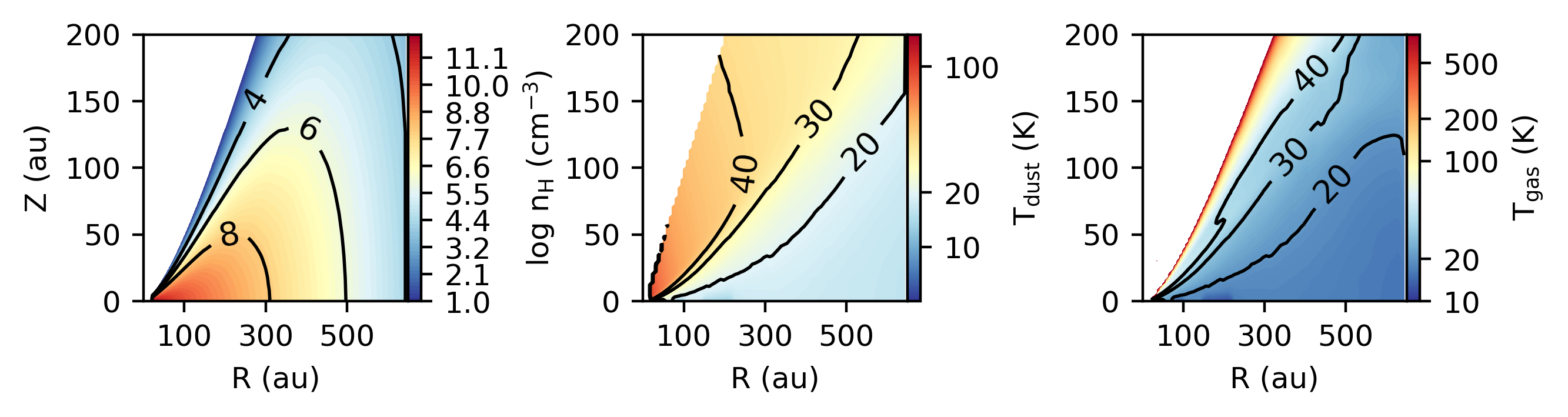}
    \caption{Maps of the H number density (left), dust temperature (center), and gas temperature (right) in our best-fit model.}
    \label{phys}
\end{figure*}

In the inner disk the model under-predicts the strength of the \co\ 2-1 emission by a factor of 2.6, while providing a reasonable fit to the emission from the less abundant CO isotopologs (Figure~\ref{profile62}). The emission surface profiles derived from observations show that \co\ emission originates from higher in the disk than the other lines. To fit the \co\ $J=2-1$ emission in the inner disk we increase the temperature inside of 32~au and for $Z/R > 0.1$. 
Increasing the gas temperature by a factor of 10 in this region, from several tens of kelvin to several hundred kelvin, greatly improves the agreement between the model \co\ $J=2-1$ emission and the observations
without significantly increasing emission from the other lines (Figure~\ref{profiles}). Possible sources of this extra heating are discussed below. Increasing the temperature in the inner disk has a negligible effect on the model HD flux. This model, with an increased temperature in the inner disk, provides the best fit to the data. The input values for our best-fit model are shown in Table~\ref{bestparams} and the 2D hydrogen gas distribution, dust temperature, and gas temperature are shown in Figure~\ref{phys}.
However, because of our sparsely sampled parameter space we cannot rule out the possibility of other model solutions fitting the data equally well.

\begin{deluxetable}{lccc}\label{bestparams}
    \tablewidth{0pt}
    \tablecolumns{2}
    \tablecaption{Gas and Dust Population Parameters: best-fit model Values}
    \tablehead{ & Gas & Small Dust & Large Dust}
     \startdata
     Mass (\msun) & 0.2 & \( 1.03\ee{-4}\) & \(5.94\ee{-4} \) \\
     \(\psi\) & 1.5 & 1.5  & 2.0\\
     \(\gamma\) & 0.59  & 0.59  & 1.0 \\
     \({\rm R}_{c}\) (au) & 111 &111 &  \nodata \\
     \({\rm R}_{ref}\) (au) & 100 & 100 &  100 \\
     \({\rm h}_{ref}\) (au) & 7.5 & 7.5 &  1.0 \\     
     \({\rm R}_{in}\) (au) & 15 & 1.0 & 0.45 \\
     \({\rm R}_{out}\) (au) & 650 & 650 & 600  \\
     \enddata
\end{deluxetable}

In order to match the \co\ 2-1 flux in the inner disk, we need to increase the gas temperature in the surface layers of the inner disk. The primary source of dust heating in \texttt{RAC2D} is radiation from the central star. The gas temperature is initially assumed to be the same as the dust temperature and allowed to evolve due to a number of heating and cooling processes, including photoelectric heating, endothermic and exothermic chemical reactions, and viscous dissipation \citep{Du14}. However, in the surface layers of the inner disk photoelectric heating of polycyclic aromatic hydrocarbons
(PAHs) can be an important contributor to the local gas temperature \citep{Kamp04,Woitke16}.
While heating by PAHs is included in \texttt{RAC2D}, the PAH abundance in the disk surface layers is uncertain. We assume a PAH abundance relative to H of 1.6\ee{-7} \citep{Du14}. A higher PAH abundance could result in increased photoelectric heating and thus in a warmer disk surface.

Alternatively, mechanical heating mechanisms such as stellar winds and accretion onto the central star can raise the temperature of the inner disk.
When including mechanical heating, both \citet{Glassgold04} and \citet{Najita17} found temperatures of several hundred kelvin for vertical column densities equivalent to those in the region of our model where we artificially increase the temperature, though these models looked at full disks without a large central dust cavity as seen in the GM Aur disk.
\citet{CalahanMAPS} also found that their \texttt{RAC2D} model requires additional heating to match the observed \co\ $J=2-1$ flux in the HD 163296 disk inside of 32 au, further suggesting that under-predicting emission in the surface layers of the inner disk is due to limitations in the code.

\begin{figure}
    \centering
    \includegraphics[width=\columnwidth]{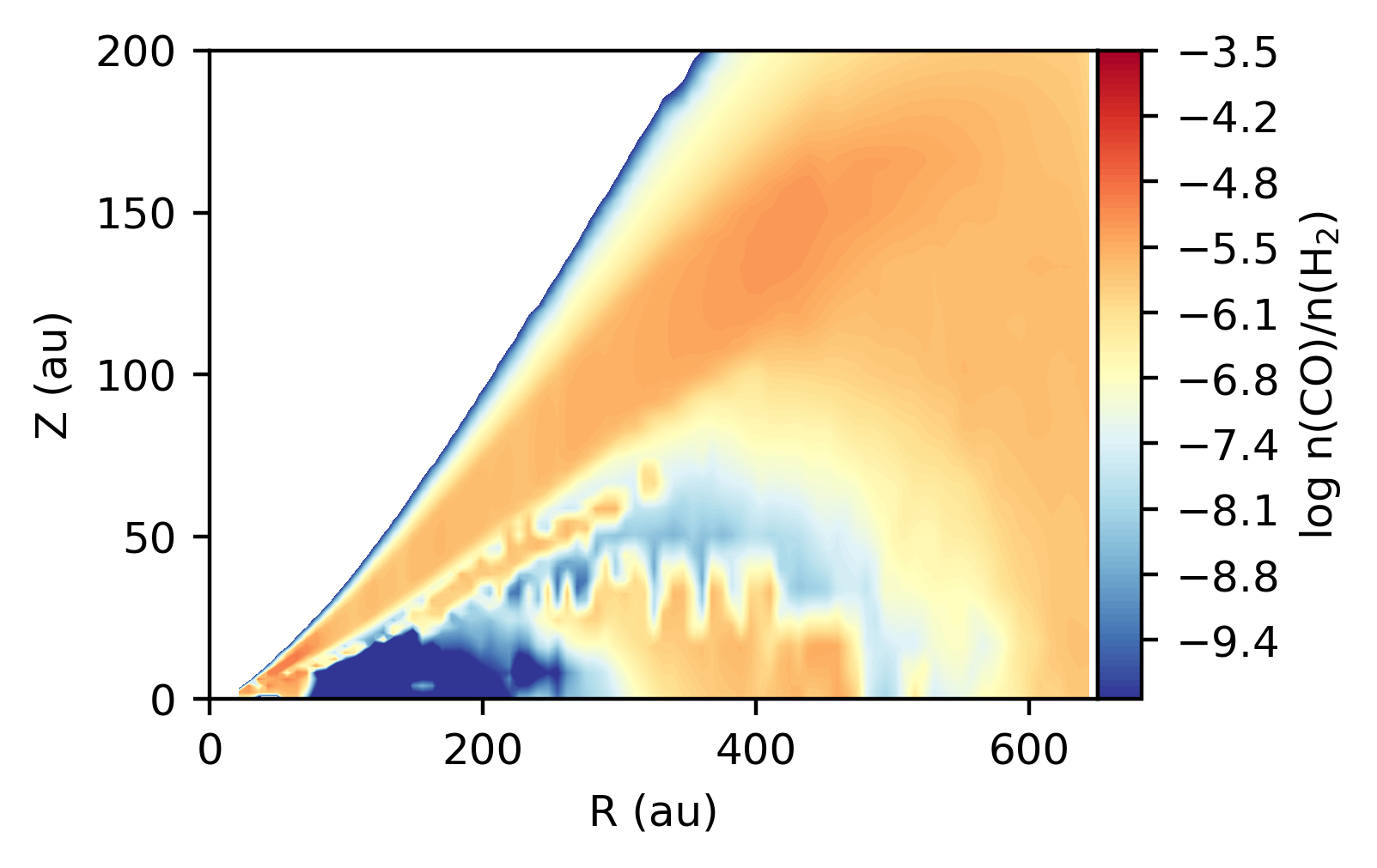}
    \caption{Map of CO abundance relative to \hh\ in our best-fit model.}
    \label{COmap}
\end{figure}

\subsection{CO Abundance}
Figure~\ref{COmap} shows the CO abundance relative to \hh\ as a function of height and radius in our best-fit model. CO is effectively frozen out from the gas near the midplane from 75 to 250 au. 
Outside the millimeter dust disk, the gas temperature remains below the CO freeze-out temperature. However, nonthermal desorption by UV photons allows some CO to remain in the gas  \citep[e.g.,][]{Oberg15}. 
In the region directly above the CO freeze-out layer, gas-phase CO has been converted into \cotwo\ ice \citep[e.g.,][]{Reboussin15,Bosman18}. Closer to the disk surface, where the temperature exceeds the \cotwo\ freeze-out temperature, CO remains in the gas. 

\begin{figure}
    \centering
    \includegraphics[width=\columnwidth]{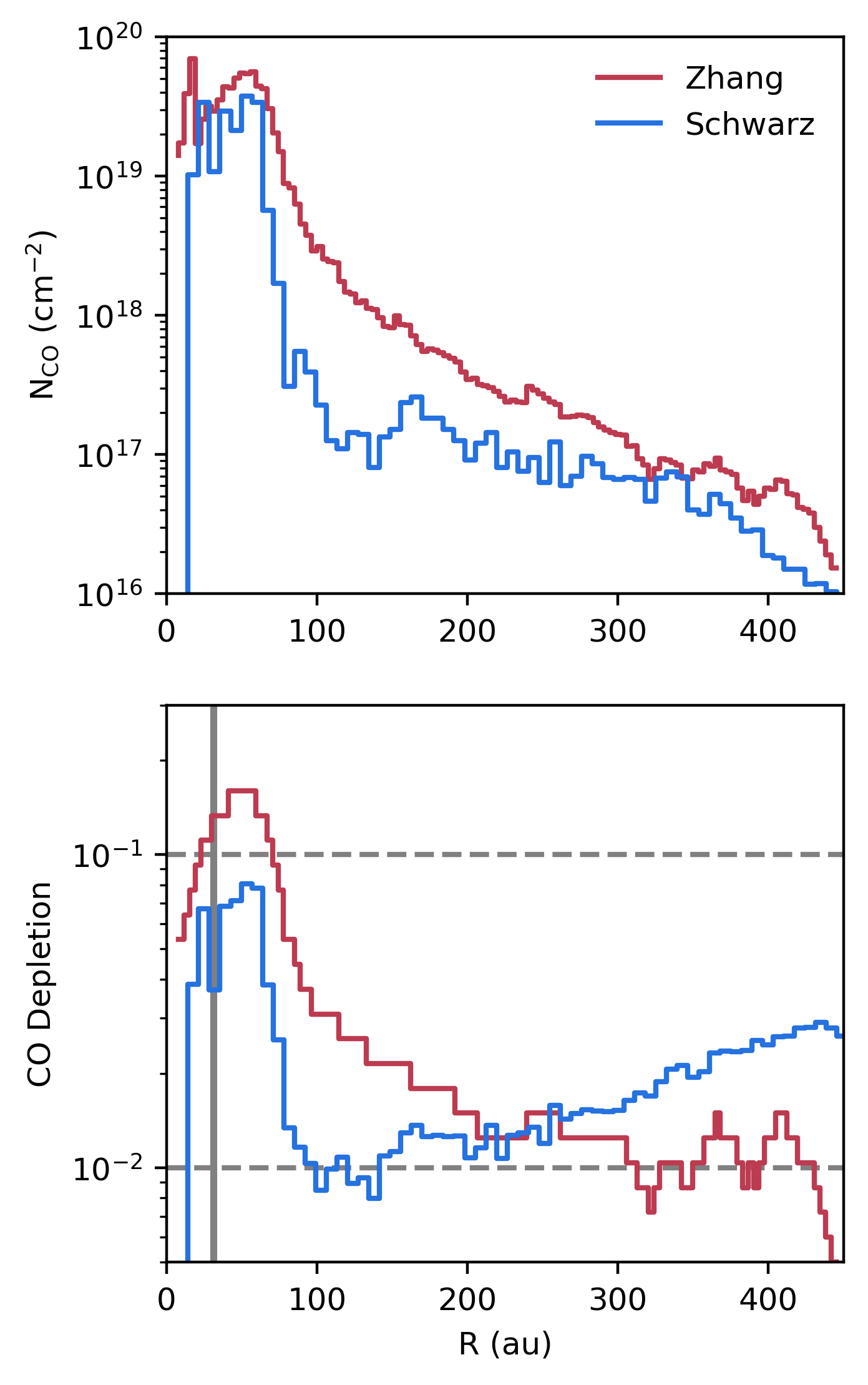}
    \caption{Top: CO column density profile for our best-fit model and for the best-fit model of \citet{ZhangMAPS}. Bottom: CO depletion factor as a function of radius for our best-fit model after evolving the chemistry for 1 Myr, as well as the best-fit model of \citet{ZhangMAPS} relative to the initial CO abundance. Vertical gray line in the midplane CO snowline at 31 au in our model. Horizontal dashed lines indicate depletion factors of 10 and 100.}
    \label{COdepprofile}
\end{figure}

Figure~\ref{COdepprofile} compares the CO depletion profile after evolving the chemistry in our best-fit model for 1 Myr to that found by \citet{ZhangMAPS}. Both results follow the general trend of a roughly constant, high level of CO depletion outside of roughly 100 au, with the inner disk less depleted in CO. The location of the midplane CO snowline in our model, here defined as where the CO gas and ice abundances are equal, is 31 au, consistent with the derived CO snowline of $30\pm5$ au from \citet{ZhangMAPS}. 
Our model has greater CO depletion inside 200 au, as well as a more abrupt return of CO in the inner disk compared to the \citet{ZhangMAPS} results.
The abrupt change in the CO column is due to the conversion of CO gas into
CO$\mathrm{_2}$ ice and CH$\mathrm{_4}$ ice near the midplane from roughly 90-150 au, 
which is not seen to the same extent in the \citet{ZhangMAPS} model.
Given that the two approaches remove CO from the disk at different points in the modeling process, and that this work attempts to match a greater number of observations, some variation is to be expected.

\section{Discussion}\label{sec:discuss}

\subsection{2D Temperature \& CO Distribution}\label{twod}
\begin{figure*}
    \centering
    \includegraphics[width=2\columnwidth]{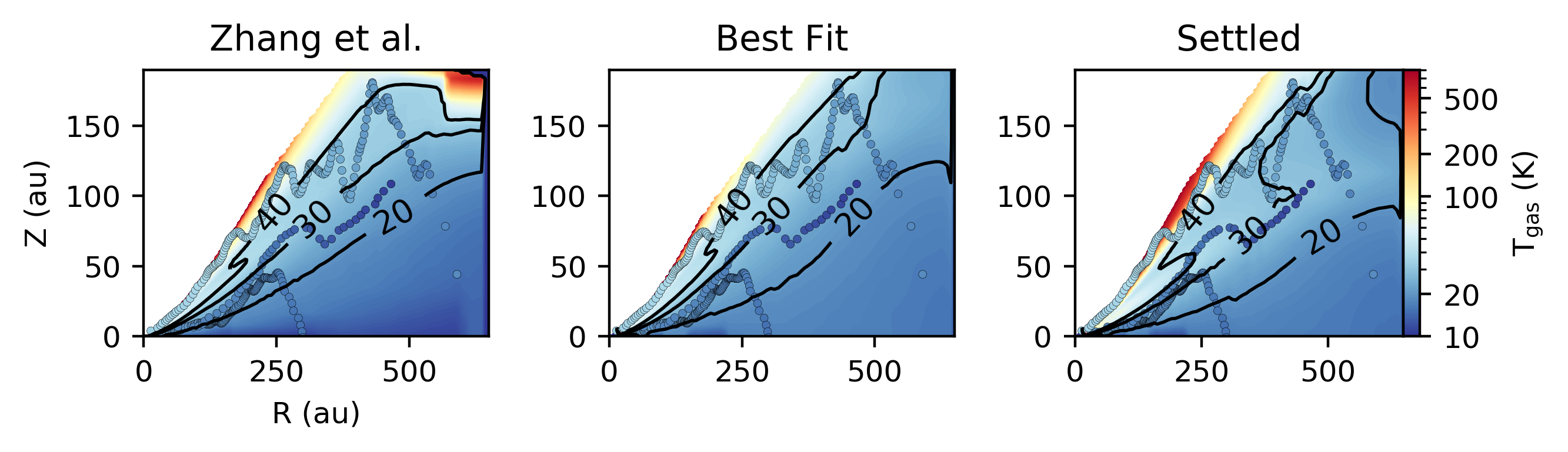}
    \caption{Temperatures of the emission surfaces derived from \co, and \coo\ 2-1 the \coo\ 3-2 observations overlaid on the gas temperature map from our best-fit model.}
    \label{Tcompare}
\end{figure*}

\begin{figure*}
    \centering
    \includegraphics[width=2\columnwidth]{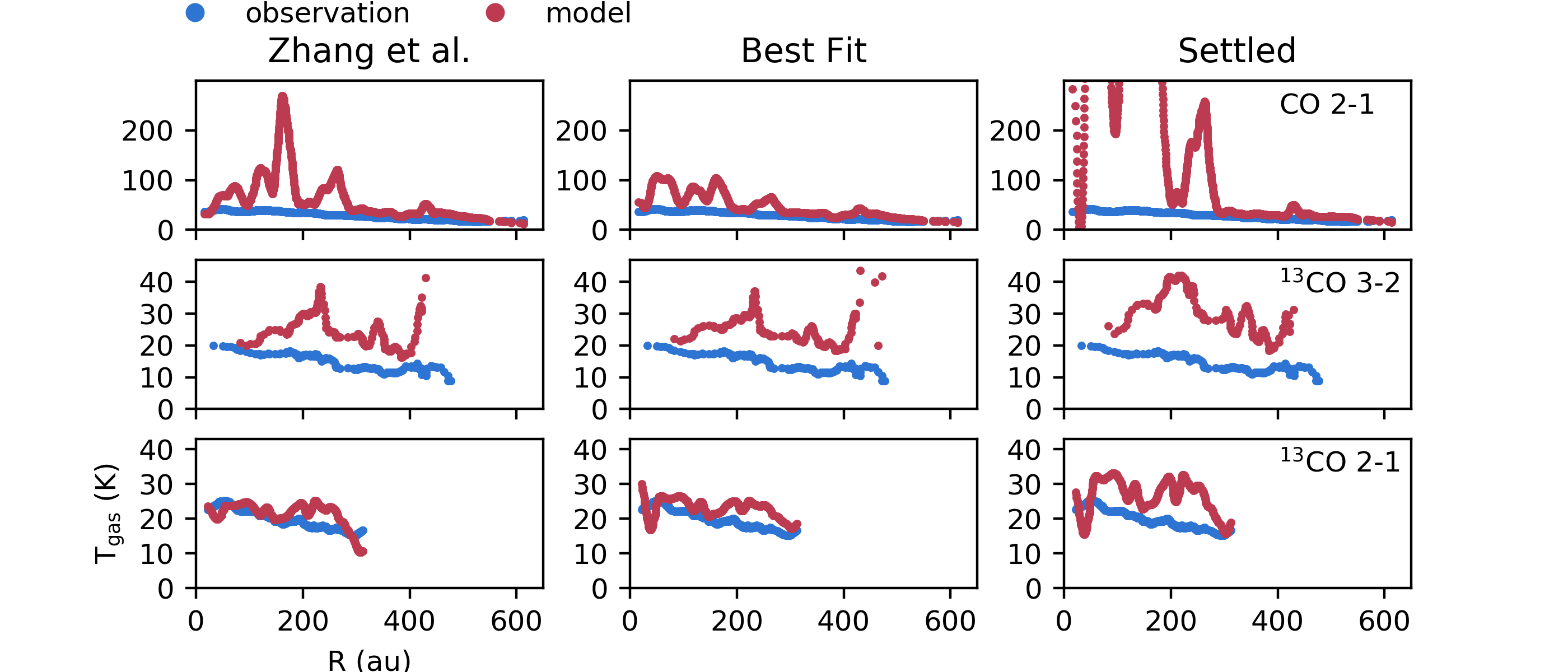}
    \caption{Temperatures of the emission surfaces derived from \co\ 2-1, \coo\ 2-1, and \coo\ 3-2 observations (blue) compared to the temperature of the corresponding location in our models (red).}
    \label{Tprofiles}
\end{figure*}

Figure~\ref{Tcompare} shows the 2D gas temperature distribution in a subset of our models with the derived temperatures for the \co\ $J=2-1$, \coo\ $J=2-1$, and \coo\ $J=3-2$ surfaces over-plotted. The \cooo\ $J=2-1$ is not included as this emission is optically thin and thus is not a good temperature tracer.
The temperature extraction follows the same process described by \citet{LawMAPS_surf}. Briefly, the gas temperature is determined from the peak surface brightness at a given radius for the non-continuum subtracted line image cubes using the full Planck function and assuming the line emission is optically thick. Using non-continuum subtracted data to measure the temperature ensures the temperature is not underestimated in the case of optically thick dust emission \citep{Weaver18}.
In the layers traced by the \coo\ $J=2-1$ and $J=3-2$ lines our best-fit model temperature is in reasonable agreement with the data, with the model gas temperature varying by less than 10 K from the measured temperature at most radii (Figure~\ref{Tprofiles}).

However, at larger heights our model is over two times warmer than the temperature measured from the \co\ $J=2-1$ emission surface. 
Past models of CO emission in protoplanetary disks have cooled the upper layers of the disk by increasing dust settling \citep{McClure16,Calahan21}. We test this solution by decreasing the scale height of the small dust grains in our model.
Decreasing the small grain scale height from 7.5 to 5 au does in fact decrease the gas temperature in some regions of the disk, particularly in the upper layers beyond 400 au (Figure~\ref{Tcompare}). However, the regions traced by CO emission are warmer in this model, increasing the discrepancy between the model and the observations (Figure~\ref{Tprofiles}).

An alternative explanation is that the upper layers of the disk are warmer and more CO rich than we assume in our models.
If material is falling onto the disk from a residual cloud or envelope, the favored explanation for the nonaxisymmetric features seen in \co\ $J=2-1$ \citep{HuangMAPS}, the infalling material is expected to produce shock heating \citep{Sakai14b}. This heating will enhance the gas temperature in the surface layers from which \co\ emits. Additionally, the infalling material is unlikely to have undergone much chemical processing, and thus will have a CO/\hh\ abundance ratio similar to that of the dense ISM. 
While the amount of CO supplied to the disk by a residual envelope is likely to be small, the combination of increased temperature and elevated CO/\hh\ could result in some \co\ emission originating closer to the disk surface.

\subsection{Mass Traced by HD \& Comparison to Previous Work}
\begin{figure}
    \centering
    \includegraphics[width=\columnwidth]{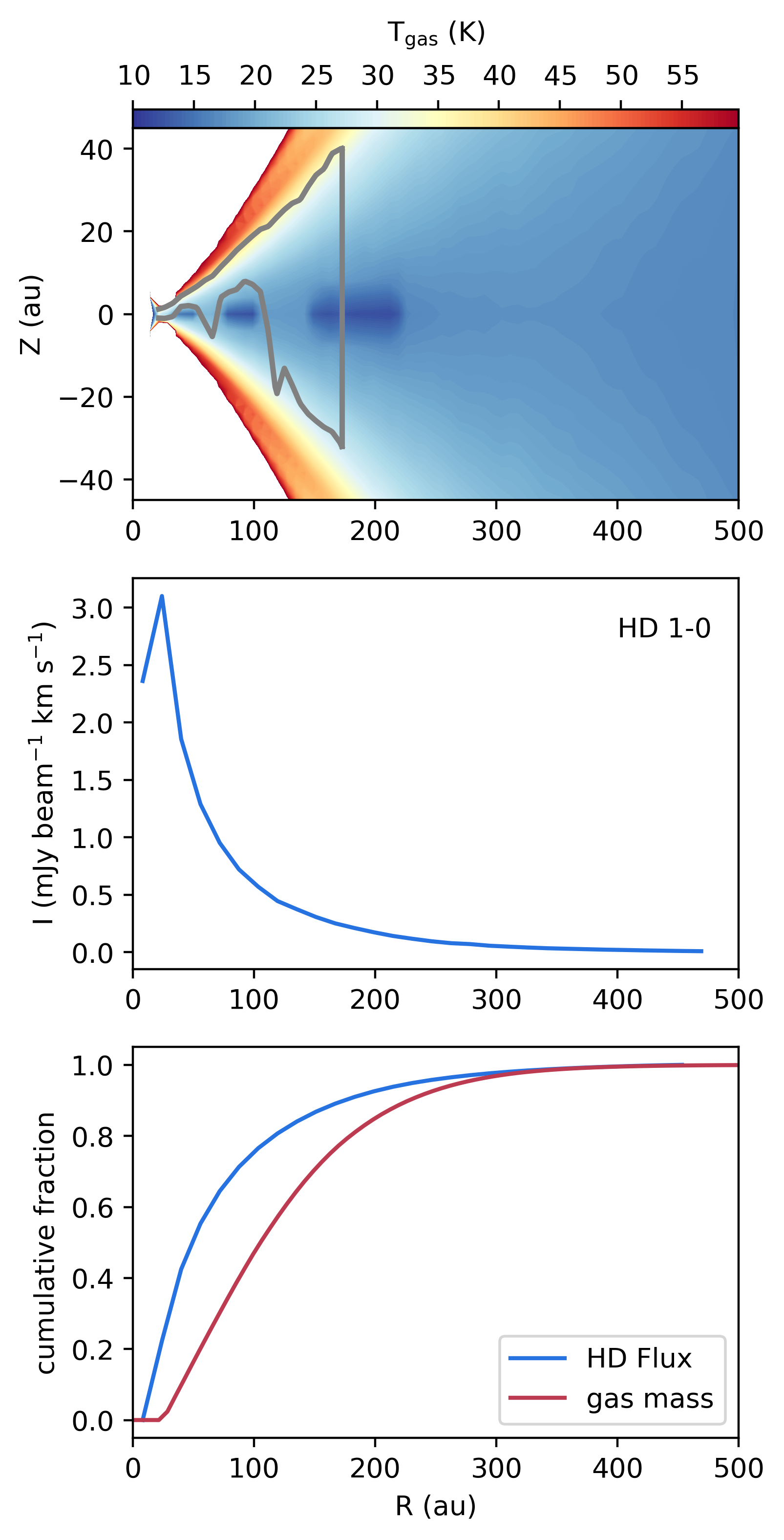}
    \caption{Top: HD $J=1-0$ emitting region in our best-fit model overlayed on a map of the gas temperature. The gray contour contains the middle 75\% of the HD emission.  Middle: Deprojected, azimuthally averaged radial profile for the HD $J=1-0$ emission. Since the observed HD emission is spatially unresolved, the model is not convolved with a beam. Bottom: Plot showing the total HD flux (blue) and gas mass (red) interior to a given radius. The HD emission preferentially originates from the warm inner disk.}
    \label{hddist}
\end{figure}

Figure~\ref{hddist} shows the distribution of the HD emission in our best-fit model. Seventy-five percent of the HD emission originates from the inner 100 au. In particular, the hot, low density gas inside the millimeter dust inner radius at 32 au contributes a non-negligible amount to the HD flux.  In comparison, only $47\%$ of the total disk gas mass is inside 100 au. Since HD does not readily emit at temperatures less than $\sim20$ K, much of the disk beyond 100 au is not well traced by the HD $J=1-0$ emission. There may be more mass in the outer disk than accounted for in our best-fit model. Additional analysis, e.g., using CS emission \citep{Teague18}, is needed to better constrain the gas density in the outer disk.

Previous analysis of the HD detection by \citet{McClure16} in GM Aur constrains the disk gas mass to 0.025-0.204 \msun. Based on analysis of the millimeter and centimeter continuum, \citet{Macias18} found a total dust mass of 2 M$\mathrm{_J}$. Assuming a gas-to-dust mass ratio of 100, this corresponds to a total gas mass of 0.19 \msun, at the high end of the range given by \citet{McClure16}. 
\citet{Trapman17} reanalyzed the HD detection in GM Aur by comparing to the HD line flux in a grid of generic 2D thermochemical models. They constrained the mass of the GM Aur disk to be between 0.01 \msun\ and a few tenths of a solar mass. 
\citet{Woitke19} have also built a model of the GM Aur disk as part of the DIANA project. Their model based on the observed SED has a total disk mass of 0.11 \msun\ and also reasonably reproduces the observed total flux of the 63 \micron\ [OI] line as well as the \co\ 2-1 and \hcop\ 3-2 lines. Their final model, independent of the SED fit and based on Submillimeter Array (SMA) observations of the \co\ $J$ = 2-1 line, has a disk mass of 3.3\ee{-2} \msun.

Our best-fit model has a total gas mass of 0.2 \msun, consistent with the upper limits of previous works. This high value comes in part from our overall low disk temperature, needed to match the gas temperature from CO observations. Additionally, previous works have set the outer radius of the disk to 250-300 au. Here we set the disk outer radius to 650 au based on the observed extent of the \co\ 2-1 emission, though less than $3\%$ of the total gas mass is outside of 300 au.

\subsection{Disk Stability}
\begin{figure}
    \centering
    \includegraphics[width=\columnwidth]{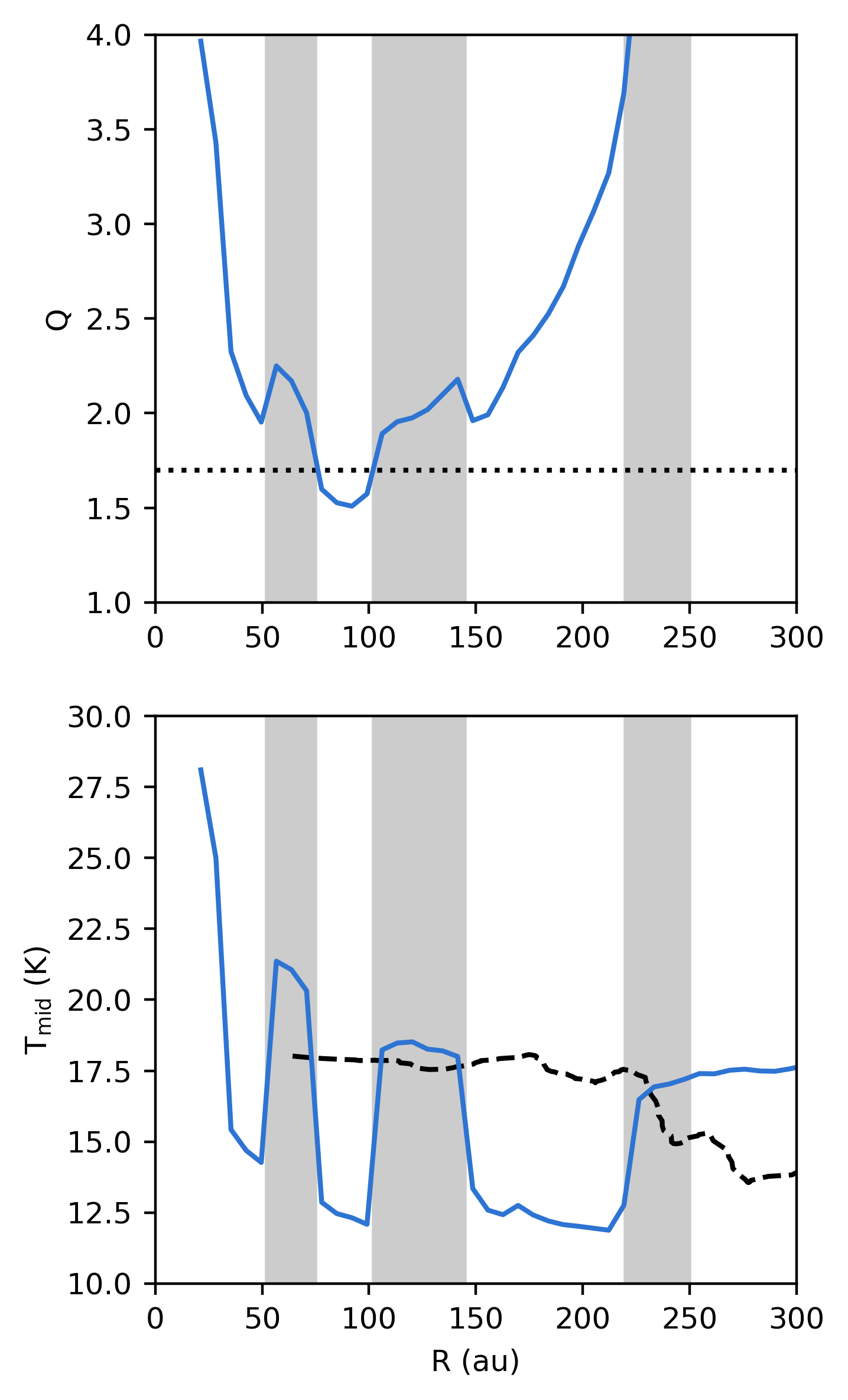}
    \caption{Top: Calculated Toomre Q value. Dotted line indicates the gravitationally unstable threshold of 1.7 for a geometrically thick disk. Bottom: Midplane temperature as a function of radius in our best-fit model. Dashed line is the gas temperature derived from the \coo\ 2-1 observations. gray regions indicate the locations of the observed gaps in the millimeter continuum \citep{Macias18}. The dips in temperature, and, by extension, $Q$, corresponds to bright rings in the continuum.}
    \label{toomreQ}
\end{figure}

The stability of a rotating disk against gravitational collapse is often characterized using the Toomre $Q$ parameter \citep{Toomre64}:
\begin{equation}
    Q = \frac{c_s \Omega}{\pi G \Sigma}
\end{equation}
where $c_s$ is the gas sound speed assuming the midplane gas temperature, $\Omega$ is the Keplerian angular velocity of the disk, and $\Sigma$ is the total gas+dust surface density. 
For a geometrically thin disk $Q \sim 1$ is needed for density perturbations to develop. However, numerical simulations demonstrate that instabilities can develop for $Q < 1.7$ in systems not well approximated by a geometrically thin disk \citep{Helled14}.
Figure~\ref{toomreQ} shows the Toomre $Q$ radial profile using values from our best fit disk model. 

Our calculated $Q$ value for GM Aur is greater than 1.7 throughout much of the disk. In the outer disk, where spiral-like features are seen in the \co\ 2-1 emission, our calculated $Q$ is extremely high, ranging from $\sim7$ at 250 au to $\sim700$ at 500 au. The two values in our model that determine $Q$ are temperature and surface density. Without changing the surface density, the midplane temperature at 250 au would need to be unrealistically low, $\sim$ 1~K, to reach a $Q$ of 1.7.

As discussed in the previous section, the HD detection does not constrain the gas surface density in the outer disk. The model surface density would need to be increased by a factor of four at 250 au to reach $Q$ of 1.7 while holding the temperature constant. \citet{ZhangMAPS} note that for four out of the five MAPS sources, including GM Aur, the CO column density profiles are very shallow, consistent with a viscously evolving disk.
This supports the conclusion of \citet{HuangMAPS}, who argue against the nonaxisymmetric features seen in \co\ $J=2-1$ being driven exclusively by disk instability based on the kinematics. 

However, $Q$ dips below 1.7 from 70 to 100 au, corresponding to the location of one of the bright rings seen in the continuum. The concentration of large dust grains in this region increases the disk opacity and thus decreases the temperature of both the gas and the dust. 
This lower temperature, in turn, results in a lower sound speed and thus in a lower Toomre $Q$ value. The presence of dust rings and gaps can lower the midplane temperature in a dust ring by several kelvin compared to a disk with a smoothly varying surface density profile \citep{Facchini18,vanderMarel18,Alarcon20,CalahanMAPS}. 


The dip in the Toomre $Q$ in our model is due entirely to an over-density of dust. Our model is able to fit the CO emission profiles in this region without a corresponding increase in the gas density.
Decreasing the disk surface density to $88\%$ of our assumed value between 70 and 100 au would bring the dust ring into a gravitationally stable regime. Alternatively, a warmer temperature than assumed would also lead to a higher $Q$.
Between 70 and 100 au our model midplane temperature is $\sim 12$ K. 
The temperature derived from the observed \coo\ 2-1 line is $\sim 22$ K and can be considered an upper limit on the midplane temperature. We use the temperature from \coo\ because the \cooo\ is optically thin at these radii and therefore not a good tracer of temperature. Taking the midplane temperature to be 22~K increases the $Q$ value to 2.0.
Interestingly, recent smoothed particle hydrodynamics modeling shows that a migrating planet can increase the local disk temperature, suppressing spiral structure and stabilizing the disk \citep{Rowther20}.


GI is thought to primarily manifest as nonaxisymmetric features. 
Previous analysis of the continuum disk emission at FUV wavelengths does not indicate any such features in the 70-120 au range \citep{Hornbeck16}. 
No nonaxisymmetric substructure is seen in the CO emission profiles in this region.
However, the intensity of the bright continuum ring at 40 au at 7 mm shows a low signal-to-noise ($\sim 2 \sigma$) asymmetry \citep{Macias18}. While the ring at 84 au is not detected at 7 mm with high enough sensitivity to enable a similar analysis, \citet{Huang20a} note that the 84 au ring is nonaxisymmetric at 1.1 mm. The ring appears wider along the major axis of the disk, which as \citet{Huang20a} demonstrate is unlikely to be an imaging artifact; instead the variation can be attributed to either nonaxisymmetric or vertical structure within the ring.



An interesting point of comparison is the HL Tau disk, which, though less evolved than GM Aur, has a similar total gas mass and a region of instability centered on a dust gap \citep{Booth20}.
Spiral structure is also seen in the \hcop\ $J$ = 3-2 emission toward HL Tau \citep{Yen19}. While this spiral structure was originally attributed to the infalling envelope, \citet{Booth20} note that the feature could also be associated with the region of instability in the disk.
Conversely, observations of \hcop\ $J$ = 3-2 toward GM Aur do not appear to deviate substantially from Keplerian rotation \citep{Huang20a}.
It is possible that the GM Aur disk is in the process of stabilizing after a period of infall or planet formation. Given the limitations of using Toomre $Q$ to determine the stability of the non-geometrically thin disk, a more detailed study of the kinematics is required to determine the stability of the GM Aur disk from 70 to 100 au.

\section{Summary and Conclusions}\label{sec:summary}
In this work we use observations of CO isotopologs in the GM Aur disk taken as part of the MAPS ALMA Large Program along with archival observations of CO from ALMA and HD from Herschel to build a model of the disk gas density and temperature structure. Based on our results we conclude the following:
\begin{itemize}
    \item Much of the disk (32\% by mass) is cooler than 20~K. As such the HD emission only traces the inner 200 au, while the gas disk extends to 650 au based on observations of \co. 
    \item We constrain the gas mass of the GM Aur disk to be $\sim 0.2$ \msun. While the total mass in the outer disk remains somewhat uncertain, only 15\% of the mass in our best-fit model is beyond 200 au. Any variation in mass in the outer disk will likely have only a small aefect on the total disk mass.
    \item The CO gas abundance relative to \hh\ is reduced by approximately one order of magnitude with respect to the ISM values inside 100 au and by two orders of magnitude outside 100 au. This is consistent with the analysis of \citet{ZhangMAPS}. Our model also shows CO gas returning to the midplane outside of the millimeter dust disk due to nonthermal desorption.
    \item Based on the calculated Toomre $Q$ parameter, the outer disk appears stable against gravitational collapse. However, $Q$ dips into the unstable regime between 70 and 100 au, corresponding to the second bright ring seen in millimeter dust emission. While there is some evidence for nonaxisymmetric features in the dust continuum at these radii, a more detailed study is needed to determine if the GM Aur disk is gravitationally unstable.
\end{itemize}

\acknowledgments
This paper makes use of the following ALMA data: ADS/JAO.ALMA\#2016.1.00565.S and  ADS/JAO.ALMA\#2018.1.01055.L. ALMA is a partnership of ESO (representing its member states), NSF (USA) and NINS (Japan), together with NRC (Canada), MOST and ASIAA (Taiwan), and KASI (Republic of Korea), in cooperation with the Republic of Chile. The Joint ALMA Observatory is operated by ESO, AUI/NRAO and NAOJ. The National Radio Astronomy Observatory is a facility of the National Science Foundation operated under cooperative agreement by Associated Universities, Inc. This work is based on observations made with Herschel, a European Space Agency Cornerstone Mission
with significant participation by NASA.

K.R.S., K.Z., J.B., J.H., and I.C. acknowledge the support of NASA through Hubble Fellowship Program grants HST-HF2-51419.001, HST-HF2-51401.001, HST-HF2-51427.001-A, HST-HF2-51460.001-A, and HST-HF2-51405.001-A awarded by the Space Telescope Science Institute, which is operated by the Association of Universities for Research in Astronomy, Inc., for NASA, under contract NAS5-26555.
J.K.C. acknowledges support from the National Aeronautics and Space Administration FINESST grant, under Grant no. 80NSSC19K1534. 
C.J.L. and J.K.C. acknowledge funding from the National Science Foundation Graduate Research Fellowship under Grant No. DGE1745303 and DGE1256260.
K.Z. acknowledges the support of the Office of the Vice Chancellor for Research and Graduate Education at the University of Wisconsin – Madison with funding from the Wisconsin Alumni Research Foundation. 
Y.A. acknowledges support by NAOJ ALMA Scientific Research Grant Code 2019-13B, and Grant-in-Aid for Scientific Research Nos. 18H05222 and 20H05847.
S.A. and J.H. acknowledge funding support from the National Aeronautics and Space Administration under Grant No. 17-XRP17 2-0012 issued through the Exoplanets Research Program.
E.A.B., A.D.B., and F.A. acknowledge support from NSF AAG Grant \#1907653.
A.S.B. acknowledges the studentship funded by the Science and Technology Facilities Council of the United Kingdom (STFC).
G.C. is supported by NAOJ ALMA Scientific Research grant Code 2019-13B.
L.I.C. gratefully acknowledges support from the David and Lucille Packard Foundation and Johnson \& Johnson's WiSTEM2D Program.
J.D.I. acknowledges support from the Science and Technology Facilities Council of the United Kingdom (STFC) under ST/T000287/1.
R.L.G. acknowledges support from a CNES research fellowship.
F.L. and R.T. acknowledge support from the Smithsonian Institution as Submillimeter Array (SMA) Fellows.
Y.L. acknowledges the financial support by the Natural Science Foundation of China (Grant No. 11973090).
F.M. acknowledges support from ANR of France under contract ANR-16-CE31-0013 (Planet-Forming-Disks)  and ANR-15-IDEX-02 (through CDP ``Origins of Life"). 
K.I\".O. acknowledges support from the Simons Foundation (SCOL \#321183) and an NSF AAG Grant (\#1907653). 
C.W. acknowledges financial support from the University of Leeds, STFC and UKRI (grant numbers ST/R000549/1, ST/T000287/1, MR/T040726/1).

\vspace{5mm}
\facilities{ALMA, Herschel}

\software{analysisUtils (\url{https://casaguides.nrao.edu/index.php/Analysis\_Utilities}),
          astropy \citep{astropy},  
          CASA \citep{McMullin07}, 
          \texttt{diskprojection}(\url{https://github.com/richteague/disksurf}),
          matplotlib \citep{matplotlib},
          numpy \citep{numpy},
          RAC2D \citep{Du14},
          RADMC-3D, \citep{Dullemond12},
          SciPy \citep{scipy},
          }
\newpage
\bibliography{allmyrefs}{}

\begin{thebibliography}{}
\expandafter\ifx\csname natexlab\endcsname\relax\def\natexlab#1{#1}\fi
\providecommand{\url}[1]{\href{#1}{#1}}
\providecommand{\dodoi}[1]{doi:~\href{http://doi.org/#1}{\nolinkurl{#1}}}
\providecommand{\doeprint}[1]{\href{http://ascl.net/#1}{\nolinkurl{http://ascl.net/#1}}}
\providecommand{\doarXiv}[1]{\href{https://arxiv.org/abs/#1}{\nolinkurl{https://arxiv.org/abs/#1}}}

\bibitem[{{Alarc{\'o}n} {et~al.}(2020){Alarc{\'o}n}, {Teague}, {Zhang},
  {Bergin}, \& {Barraza-Alfaro}}]{Alarcon20}
{Alarc{\'o}n}, F., {Teague}, R., {Zhang}, K., {Bergin}, E.~A., \&
  {Barraza-Alfaro}, M. 2020, \apj, 905, 68, \dodoi{10.3847/1538-4357/abc1d6}

\bibitem[{{ALMA Partnership} {et~al.}(2015){ALMA Partnership}, {Brogan},
  {P{\'e}rez}, {Hunter}, {Dent}, {Hales}, {Hills}, {Corder}, {Fomalont},
  {Vlahakis}, {Asaki}, {Barkats}, {Hirota}, {Hodge}, {Impellizzeri}, {Kneissl},
  {Liuzzo}, {Lucas}, {Marcelino}, {Matsushita}, {Nakanishi}, {Phillips},
  {Richards}, {Toledo}, {Aladro}, {Broguiere}, {Cortes}, {Cortes}, {Espada},
  {Galarza}, {Garcia-Appadoo}, {Guzman-Ramirez}, {Humphreys}, {Jung}, {Kameno},
  {Laing}, {Leon}, {Marconi}, {Mignano}, {Nikolic}, {Nyman}, {Radiszcz},
  {Remijan}, {Rod{\'o}n}, {Sawada}, {Takahashi}, {Tilanus}, {Vila Vilaro},
  {Watson}, {Wiklind}, {Akiyama}, {Chapillon}, {de Gregorio-Monsalvo}, {Di
  Francesco}, {Gueth}, {Kawamura}, {Lee}, {Nguyen Luong}, {Mangum}, {Pietu},
  {Sanhueza}, {Saigo}, {Takakuwa}, {Ubach}, {van Kempen}, {Wootten},
  {Castro-Carrizo}, {Francke}, {Gallardo}, {Garcia}, {Gonzalez}, {Hill},
  {Kaminski}, {Kurono}, {Liu}, {Lopez}, {Morales}, {Plarre}, {Schieven},
  {Testi}, {Videla}, {Villard}, {Andreani}, {Hibbard}, \& {Tatematsu}}]{ALMA15}
{ALMA Partnership}, {Brogan}, C.~L., {P{\'e}rez}, L.~M., {et~al.} 2015, \apjl,
  808, L3, \dodoi{10.1088/2041-8205/808/1/L3}

\bibitem[{{Andrews} {et~al.}(2011){Andrews}, {Wilner}, {Espaillat}, {Hughes},
  {Dullemond}, {McClure}, {Qi}, \& {Brown}}]{Andrews11}
{Andrews}, S.~M., {Wilner}, D.~J., {Espaillat}, C., {et~al.} 2011, \apj, 732,
  42, \dodoi{10.1088/0004-637X/732/1/42}

\bibitem[{{Ansdell} {et~al.}(2016){Ansdell}, {Williams}, {van der Marel},
  {Carpenter}, {Guidi}, {Hogerheijde}, {Mathews}, {Manara}, {Miotello},
  {Natta}, {Oliveira}, {Tazzari}, {Testi}, {van Dishoeck}, \& {van
  Terwisga}}]{Ansdell16}
{Ansdell}, M., {Williams}, J.~P., {van der Marel}, N., {et~al.} 2016, \apj,
  828, 46, \dodoi{10.3847/0004-637X/828/1/46}

\bibitem[{{Astropy Collaboration} {et~al.}(2013){Astropy Collaboration},
  {Robitaille}, {Tollerud}, {Greenfield}, {Droettboom}, {Bray}, {Aldcroft},
  {Davis}, {Ginsburg}, {Price-Whelan}, {Kerzendorf}, {Conley}, {Crighton},
  {Barbary}, {Muna}, {Ferguson}, {Grollier}, {Parikh}, {Nair}, {Unther},
  {Deil}, {Woillez}, {Conseil}, {Kramer}, {Turner}, {Singer}, {Fox}, {Weaver},
  {Zabalza}, {Edwards}, {Azalee Bostroem}, {Burke}, {Casey}, {Crawford},
  {Dencheva}, {Ely}, {Jenness}, {Labrie}, {Lim}, {Pierfederici}, {Pontzen},
  {Ptak}, {Refsdal}, {Servillat}, \& {Streicher}}]{astropy}
{Astropy Collaboration}, {Robitaille}, T.~P., {Tollerud}, E.~J., {et~al.} 2013,
  \aap, 558, A33, \dodoi{10.1051/0004-6361/201322068}

\bibitem[{{Bergin} \& {Williams}(2017)}]{Bergin17}
{Bergin}, E.~A., \& {Williams}, J.~P. 2017, {The Determination of
  Protoplanetary Disk Masses}, ed. M.~{Pessah} \& O.~{Gressel}, Vol. 445, 1,
  \dodoi{10.1007/978-3-319-60609-5\_1}

\bibitem[{{Bergin} {et~al.}(2013){Bergin}, {Cleeves}, {Gorti}, {Zhang},
  {Blake}, {Green}, {Andrews}, {Evans}, {Henning}, {{\"O}berg}, {Pontoppidan},
  {Qi}, {Salyk}, \& {van Dishoeck}}]{Bergin13}
{Bergin}, E.~A., {Cleeves}, L.~I., {Gorti}, U., {et~al.} 2013, \nat, 493, 644,
  \dodoi{10.1038/nature11805}

\bibitem[{{Birnstiel} {et~al.}(2018){Birnstiel}, {Dullemond}, {Zhu}, {Andrews},
  {Bai}, {Wilner}, {Carpenter}, {Huang}, {Isella}, {Benisty}, {P{\'e}rez}, \&
  {Zhang}}]{Birnstiel18}
{Birnstiel}, T., {Dullemond}, C.~P., {Zhu}, Z., {et~al.} 2018, \apjl, 869, L45,
  \dodoi{10.3847/2041-8213/aaf743}

\bibitem[{{Booth} \& {Ilee}(2020)}]{Booth20}
{Booth}, A.~S., \& {Ilee}, J.~D. 2020, \mnras, 493, L108,
  \dodoi{10.1093/mnrasl/slaa014}

\bibitem[{{Booth} {et~al.}(2019){Booth}, {Walsh}, {Ilee}, {Notsu}, {Qi},
  {Nomura}, \& {Akiyama}}]{Booth19}
{Booth}, A.~S., {Walsh}, C., {Ilee}, J.~D., {et~al.} 2019, \apjl, 882, L31,
  \dodoi{10.3847/2041-8213/ab3645}

\bibitem[{{Bosman} {et~al.}(2018){Bosman}, {Walsh}, \& {van
  Dishoeck}}]{Bosman18}
{Bosman}, A.~D., {Walsh}, C., \& {van Dishoeck}, E.~F. 2018, \aap, 618, A182,
  \dodoi{10.1051/0004-6361/201833497}

\bibitem[{{Boss}(1997)}]{Boss97}
{Boss}, A.~P. 1997, Science, 276, 1836, \dodoi{10.1126/science.276.5320.1836}

\bibitem[{{Calahan} \& {MAPS Team}(2021)}]{CalahanMAPS}
{Calahan}, J., \& {MAPS Team}. 2021, \apj, 0, 0, \dodoi{0}

\bibitem[{{Calahan} {et~al.}(2021){Calahan}, {Bergin}, {Zhang}, {Teague},
  {Cleeves}, {Bergner}, {Blake}, {Cazzoletti}, {Guzm{\'a}n}, {Hogerheijde},
  {Huang}, {Kama}, {Loomis}, {{\"O}berg}, {Qi}, {van Dishoeck}, {Terwisscha van
  Scheltinga}, {Walsh}, \& {Wilner}}]{Calahan21}
{Calahan}, J.~K., {Bergin}, E., {Zhang}, K., {et~al.} 2021, \apj, 908, 8,
  \dodoi{10.3847/1538-4357/abd255}

\bibitem[{{Calvet} {et~al.}(2005){Calvet}, {D'Alessio}, {Watson},
  {Franco-Hern{\'a}ndez}, {Furlan}, {Green}, {Sutter}, {Forrest}, {Hartmann},
  {Uchida}, {Keller}, {Sargent}, {Najita}, {Herter}, {Barry}, \&
  {Hall}}]{Calvet05}
{Calvet}, N., {D'Alessio}, P., {Watson}, D.~M., {et~al.} 2005, \apjl, 630,
  L185, \dodoi{10.1086/491652}

\bibitem[{{Cleeves} {et~al.}(2013){Cleeves}, {Adams}, \& {Bergin}}]{Cleeves13}
{Cleeves}, L.~I., {Adams}, F.~C., \& {Bergin}, E.~A. 2013, \apj, 772, 5,
  \dodoi{10.1088/0004-637X/772/1/5}

\bibitem[{{Czekala} \& {MAPS Team}(2021)}]{CzekalaMAPS}
{Czekala}, I., \& {MAPS Team}, A. 2021, \apj, 0, 0, \dodoi{0}

\bibitem[{{Du} \& {Bergin}(2014)}]{Du14}
{Du}, F., \& {Bergin}, E.~A. 2014, \apj, 792, 2,
  \dodoi{10.1088/0004-637X/792/1/2}

\bibitem[{{Dullemond} {et~al.}(2012){Dullemond}, {Juhasz}, {Pohl}, {Sereshti},
  {Shetty}, {Peters}, {Commercon}, \& {Flock}}]{Dullemond12}
{Dullemond}, C.~P., {Juhasz}, A., {Pohl}, A., {et~al.} 2012, {RADMC-3D: A
  multi-purpose radiative transfer tool}.
\newblock \doeprint{1202.015}

\bibitem[{{Eistrup} {et~al.}(2018){Eistrup}, {Walsh}, \& {van
  Dishoeck}}]{Eistrup18}
{Eistrup}, C., {Walsh}, C., \& {van Dishoeck}, E.~F. 2018, \aap, 613, A14,
  \dodoi{10.1051/0004-6361/201731302}

\bibitem[{{Espaillat} {et~al.}(2011){Espaillat}, {Furlan}, {D'Alessio},
  {Sargent}, {Nagel}, {Calvet}, {Watson}, \& {Muzerolle}}]{Espaillat11}
{Espaillat}, C., {Furlan}, E., {D'Alessio}, P., {et~al.} 2011, \apj, 728, 49,
  \dodoi{10.1088/0004-637X/728/1/49}

\bibitem[{{Facchini} {et~al.}(2018){Facchini}, {Pinilla}, {van Dishoeck}, \&
  {de Juan Ovelar}}]{Facchini18}
{Facchini}, S., {Pinilla}, P., {van Dishoeck}, E.~F., \& {de Juan Ovelar}, M.
  2018, \aap, 612, A104, \dodoi{10.1051/0004-6361/201731390}

\bibitem[{{Facchini} {et~al.}(2019){Facchini}, {van Dishoeck}, {Manara},
  {Tazzari}, {Maud}, {Cazzoletti}, {Rosotti}, {van der Marel}, {Pinilla}, \&
  {Clarke}}]{Facchini19}
{Facchini}, S., {van Dishoeck}, E.~F., {Manara}, C.~F., {et~al.} 2019, \aap,
  626, L2, \dodoi{10.1051/0004-6361/201935496}

\bibitem[{{Favre} {et~al.}(2013){Favre}, {Cleeves}, {Bergin}, {Qi}, \&
  {Blake}}]{Favre13}
{Favre}, C., {Cleeves}, L.~I., {Bergin}, E.~A., {Qi}, C., \& {Blake}, G.~A.
  2013, \apjl, 776, L38, \dodoi{10.1088/2041-8205/776/2/L38}

\bibitem[{{Gaia Collaboration} {et~al.}(2018){Gaia Collaboration}, {Brown},
  {Vallenari}, {Prusti}, {de Bruijne}, {Babusiaux}, {Bailer-Jones}, {Biermann},
  {Evans}, {Eyer}, {Jansen}, {Jordi}, {Klioner}, {Lammers}, {Lindegren},
  {Luri}, {Mignard}, {Panem}, {Pourbaix}, {Randich}, {Sartoretti}, {Siddiqui},
  {Soubiran}, {van Leeuwen}, {Walton}, {Arenou}, {Bastian}, {Cropper},
  {Drimmel}, {Katz}, {Lattanzi}, {Bakker}, {Cacciari}, {Casta{\~n}eda},
  {Chaoul}, {Cheek}, {De Angeli}, {Fabricius}, {Guerra}, {Holl}, {Masana},
  {Messineo}, {Mowlavi}, {Nienartowicz}, {Panuzzo}, {Portell}, {Riello},
  {Seabroke}, {Tanga}, {Th{\'e}venin}, {Gracia-Abril}, {Comoretto},
  {Garcia-Reinaldos}, {Teyssier}, {Altmann}, {Andrae}, {Audard},
  {Bellas-Velidis}, {Benson}, {Berthier}, {Blomme}, {Burgess}, {Busso},
  {Carry}, {Cellino}, {Clementini}, {Clotet}, {Creevey}, {Davidson}, {De
  Ridder}, {Delchambre}, {Dell'Oro}, {Ducourant},
  {Fern{\'a}ndez-Hern{\'a}ndez}, {Fouesneau}, {Fr{\'e}mat}, {Galluccio},
  {Garc{\'\i}a-Torres}, {Gonz{\'a}lez-N{\'u}{\~n}ez}, {Gonz{\'a}lez-Vidal},
  {Gosset}, {Guy}, {Halbwachs}, {Hambly}, {Harrison}, {Hern{\'a}ndez},
  {Hestroffer}, {Hodgkin}, {Hutton}, {Jasniewicz}, {Jean-Antoine-Piccolo},
  {Jordan}, {Korn}, {Krone-Martins}, {Lanzafame}, {Lebzelter}, {L{\"o}ffler},
  {Manteiga}, {Marrese}, {Mart{\'\i}n-Fleitas}, {Moitinho}, {Mora}, {Muinonen},
  {Osinde}, {Pancino}, {Pauwels}, {Petit}, {Recio-Blanco}, {Richards},
  {Rimoldini}, {Robin}, {Sarro}, {Siopis}, {Smith}, {Sozzetti}, {S{\"u}veges},
  {Torra}, {van Reeven}, {Abbas}, {Abreu Aramburu}, {Accart}, {Aerts},
  {Altavilla}, {{\'A}lvarez}, {Alvarez}, {Alves}, {Anderson}, {Andrei},
  {Anglada Varela}, {Antiche}, {Antoja}, {Arcay}, {Astraatmadja}, {Bach},
  {Baker}, {Balaguer-N{\'u}{\~n}ez}, {Balm}, {Barache}, {Barata}, {Barbato},
  {Barblan}, {Barklem}, {Barrado}, {Barros}, {Barstow}, {Bartholom{\'e}
  Mu{\~n}oz}, {Bassilana}, {Becciani}, {Bellazzini}, {Berihuete}, {Bertone},
  {Bianchi}, {Bienaym{\'e}}, {Blanco-Cuaresma}, {Boch}, {Boeche}, {Bombrun},
  {Borrachero}, {Bossini}, {Bouquillon}, {Bourda}, {Bragaglia}, {Bramante},
  {Breddels}, {Bressan}, {Brouillet}, {Br{\"u}semeister}, {Brugaletta},
  {Bucciarelli}, {Burlacu}, {Busonero}, {Butkevich}, {Buzzi}, {Caffau},
  {Cancelliere}, {Cannizzaro}, {Cantat-Gaudin}, {Carballo}, {Carlucci},
  {Carrasco}, {Casamiquela}, {Castellani}, {Castro-Ginard}, {Charlot},
  {Chemin}, {Chiavassa}, {Cocozza}, {Costigan}, {Cowell}, {Crifo}, {Crosta},
  {Crowley}, {Cuypers}, {Dafonte}, {Damerdji}, {Dapergolas}, {David}, {David},
  {de Laverny}, {De Luise}, {De March}, {de Martino}, {de Souza}, {de Torres},
  {Debosscher}, {del Pozo}, {Delbo}, {Delgado}, {Delgado}, {Di Matteo},
  {Diakite}, {Diener}, {Distefano}, {Dolding}, {Drazinos}, {Dur{\'a}n},
  {Edvardsson}, {Enke}, {Eriksson}, {Esquej}, {Eynard Bontemps}, {Fabre},
  {Fabrizio}, {Faigler}, {Falc{\~a}o}, {Farr{\`a}s Casas}, {Federici},
  {Fedorets}, {Fernique}, {Figueras}, {Filippi}, {Findeisen}, {Fonti},
  {Fraile}, {Fraser}, {Fr{\'e}zouls}, {Gai}, {Galleti}, {Garabato},
  {Garc{\'\i}a-Sedano}, {Garofalo}, {Garralda}, {Gavel}, {Gavras}, {Gerssen},
  {Geyer}, {Giacobbe}, {Gilmore}, {Girona}, {Giuffrida}, {Glass}, {Gomes},
  {Granvik}, {Gueguen}, {Guerrier}, {Guiraud}, {Guti{\'e}rrez-S{\'a}nchez},
  {Haigron}, {Hatzidimitriou}, {Hauser}, {Haywood}, {Heiter}, {Helmi}, {Heu},
  {Hilger}, {Hobbs}, {Hofmann}, {Holland}, {Huckle}, {Hypki}, {Icardi},
  {Jan{\ss}en}, {Jevardat de Fombelle}, {Jonker}, {Juh{\'a}sz}, {Julbe},
  {Karampelas}, {Kewley}, {Klar}, {Kochoska}, {Kohley}, {Kolenberg},
  {Kontizas}, {Kontizas}, {Koposov}, {Kordopatis}, {Kostrzewa-Rutkowska},
  {Koubsky}, {Lambert}, {Lanza}, {Lasne}, {Lavigne}, {Le Fustec}, {Le
  Poncin-Lafitte}, {Lebreton}, {Leccia}, {Leclerc}, {Lecoeur-Taibi},
  {Lenhardt}, {Leroux}, {Liao}, {Licata}, {Lindstr{\o}m}, {Lister}, {Livanou},
  {Lobel}, {L{\'o}pez}, {Managau}, {Mann}, {Mantelet}, {Marchal}, {Marchant},
  {Marconi}, {Marinoni}, {Marschalk{\'o}}, {Marshall}, {Martino}, {Marton},
  {Mary}, {Massari}, {Matijevi{\v{c}}}, {Mazeh}, {McMillan}, {Messina},
  {Michalik}, {Millar}, {Molina}, {Molinaro}, {Moln{\'a}r}, {Montegriffo},
  {Mor}, {Morbidelli}, {Morel}, {Morris}, {Mulone}, {Muraveva}, {Musella},
  {Nelemans}, {Nicastro}, {Noval}, {O'Mullane}, {Ord{\'e}novic},
  {Ord{\'o}{\~n}ez-Blanco}, {Osborne}, {Pagani}, {Pagano}, {Pailler},
  {Palacin}, {Palaversa}, {Panahi}, {Pawlak}, {Piersimoni}, {Pineau}, {Plachy},
  {Plum}, {Poggio}, {Poujoulet}, {Pr{\v{s}}a}, {Pulone}, {Racero}, {Ragaini},
  {Rambaux}, {Ramos-Lerate}, {Regibo}, {Reyl{\'e}}, {Riclet}, {Ripepi}, {Riva},
  {Rivard}, {Rixon}, {Roegiers}, {Roelens}, {Romero-G{\'o}mez}, {Rowell},
  {Royer}, {Ruiz-Dern}, {Sadowski}, {Sagrist{\`a} Sell{\'e}s}, {Sahlmann},
  {Salgado}, {Salguero}, {Sanna}, {Santana-Ros}, {Sarasso}, {Savietto},
  {Schultheis}, {Sciacca}, {Segol}, {Segovia}, {S{\'e}gransan}, {Shih},
  {Siltala}, {Silva}, {Smart}, {Smith}, {Solano}, {Solitro}, {Sordo}, {Soria
  Nieto}, {Souchay}, {Spagna}, {Spoto}, {Stampa}, {Steele},
  {Steidelm{\"u}ller}, {Stephenson}, {Stoev}, {Suess}, {Surdej}, {Szabados},
  {Szegedi-Elek}, {Tapiador}, {Taris}, {Tauran}, {Taylor}, {Teixeira},
  {Terrett}, {Teyssandier}, {Thuillot}, {Titarenko}, {Torra Clotet}, {Turon},
  {Ulla}, {Utrilla}, {Uzzi}, {Vaillant}, {Valentini}, {Valette}, {van Elteren},
  {Van Hemelryck}, {van Leeuwen}, {Vaschetto}, {Vecchiato}, {Veljanoski},
  {Viala}, {Vicente}, {Vogt}, {von Essen}, {Voss}, {Votruba}, {Voutsinas},
  {Walmsley}, {Weiler}, {Wertz}, {Wevers}, {Wyrzykowski}, {Yoldas},
  {{\v{Z}}erjal}, {Ziaeepour}, {Zorec}, {Zschocke}, {Zucker}, {Zurbach}, \&
  {Zwitter}}]{Gaia18}
{Gaia Collaboration}, {Brown}, A.~G.~A., {Vallenari}, A., {et~al.} 2018, \aap,
  616, A1, \dodoi{10.1051/0004-6361/201833051}

\bibitem[{{Glassgold} {et~al.}(2004){Glassgold}, {Najita}, \&
  {Igea}}]{Glassgold04}
{Glassgold}, A.~E., {Najita}, J., \& {Igea}, J. 2004, \apj, 615, 972,
  \dodoi{10.1086/424509}

\bibitem[{{Guilloteau} \& {Dutrey}(1998)}]{Guilloteau98}
{Guilloteau}, S., \& {Dutrey}, A. 1998, \aap, 339, 467

\bibitem[{{Hasegawa} {et~al.}(1992){Hasegawa}, {Herbst}, \&
  {Leung}}]{Hasegawa92}
{Hasegawa}, T.~I., {Herbst}, E., \& {Leung}, C.~M. 1992, \apjs, 82, 167,
  \dodoi{10.1086/191713}

\bibitem[{{Helled} {et~al.}(2014){Helled}, {Bodenheimer}, {Podolak}, {Boley},
  {Meru}, {Nayakshin}, {Fortney}, {Mayer}, {Alibert}, \& {Boss}}]{Helled14}
{Helled}, R., {Bodenheimer}, P., {Podolak}, M., {et~al.} 2014, in Protostars
  and Planets VI, ed. H.~{Beuther}, R.~S. {Klessen}, C.~P. {Dullemond}, \&
  T.~{Henning}, 643, \dodoi{10.2458/azu\_uapress\_9780816531240-ch028}

\bibitem[{{Hogerheijde} {et~al.}(2011){Hogerheijde}, {Bergin}, {Brinch},
  {Cleeves}, {Fogel}, {Blake}, {Dominik}, {Lis}, {Melnick}, {Neufeld},
  {Pani{\'c}}, {Pearson}, {Kristensen}, {Y{\i}ld{\i}z}, \& {van
  Dishoeck}}]{Hogerheijde11}
{Hogerheijde}, M.~R., {Bergin}, E.~A., {Brinch}, C., {et~al.} 2011, Science,
  334, 338, \dodoi{10.1126/science.1208931}

\bibitem[{{Hornbeck} {et~al.}(2016){Hornbeck}, {Swearingen}, {Grady},
  {Williger}, {Brown}, {Sitko}, {Wisniewski}, {Perrin}, {Lauroesch},
  {Schneider}, {Apai}, {Brittain}, {Brown}, {Champney}, {Hamaguchi}, {Henning},
  {Lynch}, {Petre}, {Russell}, {Walter}, \& {Woodgate}}]{Hornbeck16}
{Hornbeck}, J.~B., {Swearingen}, J.~R., {Grady}, C.~A., {et~al.} 2016, \apj,
  829, 65, \dodoi{10.3847/0004-637X/829/2/65}

\bibitem[{{Huang} \& {MAPS Team}(2021)}]{HuangMAPS}
{Huang}, J., \& {MAPS Team}. 2021, \apj, 0, 0, \dodoi{0}

\bibitem[{{Huang} {et~al.}(2020){Huang}, {Andrews}, {Dullemond}, {{\"O}berg},
  {Qi}, {Zhu}, {Birnstiel}, {Carpenter}, {Isella}, {Mac{\'\i}as}, {McClure},
  {P{\'e}rez}, {Teague}, {Wilner}, \& {Zhang}}]{Huang20a}
{Huang}, J., {Andrews}, S.~M., {Dullemond}, C.~P., {et~al.} 2020, \apj, 891,
  48, \dodoi{10.3847/1538-4357/ab711e}

\bibitem[{{Hughes} {et~al.}(2009){Hughes}, {Andrews}, {Espaillat}, {Wilner},
  {Calvet}, {D'Alessio}, {Qi}, {Williams}, \& {Hogerheijde}}]{Hughes09}
{Hughes}, A.~M., {Andrews}, S.~M., {Espaillat}, C., {et~al.} 2009, \apj, 698,
  131, \dodoi{10.1088/0004-637X/698/1/131}

\bibitem[{{Hunter}(2007)}]{matplotlib}
{Hunter}, J.~D. 2007, Computing in Science and Engineering, 9, 90,
  \dodoi{10.1109/MCSE.2007.55}

\bibitem[{{Isella} {et~al.}(2016){Isella}, {Guidi}, {Testi}, {Liu}, {Li}, {Li},
  {Weaver}, {Boehler}, {Carperter}, {De Gregorio-Monsalvo}, {Manara}, {Natta},
  {P{\'e}rez}, {Ricci}, {Sargent}, {Tazzari}, \& {Turner}}]{Isella16}
{Isella}, A., {Guidi}, G., {Testi}, L., {et~al.} 2016, \prl, 117, 251101,
  \dodoi{10.1103/PhysRevLett.117.251101}

\bibitem[{Jones {et~al.}(2001--)Jones, Oliphant, Peterson, {et~al.}}]{scipy}
Jones, E., Oliphant, T., Peterson, P., {et~al.} 2001--, {SciPy}: Open source
  scientific tools for {Python}.
\newblock \url{http://www.scipy.org/}

\bibitem[{{Kama} {et~al.}(2020){Kama}, {Trapman}, {Fedele}, {Bruderer},
  {Hogerheijde}, {Miotello}, {van Dishoeck}, {Clarke}, \& {Bergin}}]{Kama20}
{Kama}, M., {Trapman}, L., {Fedele}, D., {et~al.} 2020, \aap, 634, A88,
  \dodoi{10.1051/0004-6361/201937124}

\bibitem[{{Kamp} \& {Dullemond}(2004)}]{Kamp04}
{Kamp}, I., \& {Dullemond}, C.~P. 2004, \apj, 615, 991, \dodoi{10.1086/424703}

\bibitem[{{Krijt} {et~al.}(2018){Krijt}, {Schwarz}, {Bergin}, \&
  {Ciesla}}]{Krijt18}
{Krijt}, S., {Schwarz}, K.~R., {Bergin}, E.~A., \& {Ciesla}, F.~J. 2018, \apj,
  864, 78, \dodoi{10.3847/1538-4357/aad69b}

\bibitem[{{Law} \& {MAPS Team}(2021)}]{LawMAPS_surf}
{Law}, C., \& {MAPS Team}. 2021, \apj, 0, 0, \dodoi{0}

\bibitem[{{Linsky}(1998)}]{Linsky98}
{Linsky}, J.~L. 1998, \ssr, 84, 285

\bibitem[{{Liu} {et~al.}(2018){Liu}, {Jin}, {Li}, {Isella}, \& {Li}}]{Liu18}
{Liu}, S.-F., {Jin}, S., {Li}, S., {Isella}, A., \& {Li}, H. 2018, \apj, 857,
  87, \dodoi{10.3847/1538-4357/aab718}

\bibitem[{{Long} {et~al.}(2017){Long}, {Herczeg}, {Pascucci}, {Drabek-Maunder},
  {Mohanty}, {Testi}, {Apai}, {Hendler}, {Henning}, {Manara}, \&
  {Mulders}}]{Long17}
{Long}, F., {Herczeg}, G.~J., {Pascucci}, I., {et~al.} 2017, \apj, 844, 99,
  \dodoi{10.3847/1538-4357/aa78fc}

\bibitem[{{Lynden-Bell} \& {Pringle}(1974)}]{LyndenBell74}
{Lynden-Bell}, D., \& {Pringle}, J.~E. 1974, \mnras, 168, 603,
  \dodoi{10.1093/mnras/168.3.603}

\bibitem[{{Mac{\'\i}as} {et~al.}(2018){Mac{\'\i}as}, {Espaillat}, {Ribas},
  {Schwarz}, {Anglada}, {Osorio}, {Carrasco-Gonz{\'a}lez}, {G{\'o}mez}, \&
  {Robinson}}]{Macias18}
{Mac{\'\i}as}, E., {Espaillat}, C.~C., {Ribas}, {\'A}., {et~al.} 2018, \apj,
  865, 37, \dodoi{10.3847/1538-4357/aad811}

\bibitem[{{Mathis} {et~al.}(1977){Mathis}, {Rumpl}, \& {Nordsieck}}]{Mathis77}
{Mathis}, J.~S., {Rumpl}, W., \& {Nordsieck}, K.~H. 1977, \apj, 217, 425,
  \dodoi{10.1086/155591}

\bibitem[{{McClure} {et~al.}(2016){McClure}, {Bergin}, {Cleeves}, {van
  Dishoeck}, {Blake}, {Evans}, {Green}, {Henning}, {{\"O}berg}, {Pontoppidan},
  \& {Salyk}}]{McClure16}
{McClure}, M.~K., {Bergin}, E.~A., {Cleeves}, L.~I., {et~al.} 2016, \apj, 831,
  167, \dodoi{10.3847/0004-637X/831/2/167}

\bibitem[{{McMullin} {et~al.}(2007){McMullin}, {Waters}, {Schiebel}, {Young},
  \& {Golap}}]{McMullin07}
{McMullin}, J.~P., {Waters}, B., {Schiebel}, D., {Young}, W., \& {Golap}, K.
  2007, in Astronomical Society of the Pacific Conference Series, Vol. 376,
  Astronomical Data Analysis Software and Systems XVI, ed. R.~A. {Shaw},
  F.~{Hill}, \& D.~J. {Bell}, 127

\bibitem[{{Miotello} {et~al.}(2014){Miotello}, {Bruderer}, \& {van
  Dishoeck}}]{Miotello14}
{Miotello}, A., {Bruderer}, S., \& {van Dishoeck}, E.~F. 2014, \aap, 572, A96,
  \dodoi{10.1051/0004-6361/201424712}

\bibitem[{{Najita} \& {{\'A}d{\'a}mkovics}(2017)}]{Najita17}
{Najita}, J.~R., \& {{\'A}d{\'a}mkovics}, M. 2017, \apj, 847, 6,
  \dodoi{10.3847/1538-4357/aa8632}

\bibitem[{{Najita} \& {Bergin}(2018)}]{Najita18}
{Najita}, J.~R., \& {Bergin}, E.~A. 2018, \apj, 864, 168,
  \dodoi{10.3847/1538-4357/aad80c}

\bibitem[{{{\"O}berg} {et~al.}(2015){{\"O}berg}, {Furuya}, {Loomis}, {Aikawa},
  {Andrews}, {Qi}, {van Dishoeck}, \& {Wilner}}]{Oberg15}
{{\"O}berg}, K.~I., {Furuya}, K., {Loomis}, R., {et~al.} 2015, \apj, 810, 112,
  \dodoi{10.1088/0004-637X/810/2/112}

\bibitem[{{{\"O}berg} \& {MAPS Team}(2021)}]{ObergMAPS}
{{\"O}berg}, K.~I., \& {MAPS Team}. 2021, \apj, 0, 0, \dodoi{0}

\bibitem[{{Pinte} {et~al.}(2018){Pinte}, {Price}, {M{\'e}nard}, {Duch{\^e}ne},
  {Dent}, {Hill}, {de Gregorio-Monsalvo}, {Hales}, \& {Mentiplay}}]{Pinte18}
{Pinte}, C., {Price}, D.~J., {M{\'e}nard}, F., {et~al.} 2018, \apj, 860, L13,
  \dodoi{10.3847/2041-8213/aac6dc}

\bibitem[{{Powell} {et~al.}(2019){Powell}, {Murray-Clay}, {P{\'e}rez},
  {Schlichting}, \& {Rosenthal}}]{Powell19}
{Powell}, D., {Murray-Clay}, R., {P{\'e}rez}, L.~M., {Schlichting}, H.~E., \&
  {Rosenthal}, M. 2019, \apj, 878, 116, \dodoi{10.3847/1538-4357/ab20ce}

\bibitem[{{Powell} {et~al.}(2017){Powell}, {Murray-Clay}, \&
  {Schlichting}}]{Powell17}
{Powell}, D., {Murray-Clay}, R., \& {Schlichting}, H.~E. 2017, \apj, 840, 93,
  \dodoi{10.3847/1538-4357/aa6d7c}

\bibitem[{{Reboussin} {et~al.}(2015){Reboussin}, {Wakelam}, {Guilloteau},
  {Hersant}, \& {Dutrey}}]{Reboussin15}
{Reboussin}, L., {Wakelam}, V., {Guilloteau}, S., {Hersant}, F., \& {Dutrey},
  A. 2015, \aap, 579, A82, \dodoi{10.1051/0004-6361/201525885}

\bibitem[{{Rowther} {et~al.}(2020){Rowther}, {Meru}, {Kennedy}, {Nealon}, \&
  {Pinte}}]{Rowther20}
{Rowther}, S., {Meru}, F., {Kennedy}, G.~M., {Nealon}, R., \& {Pinte}, C. 2020,
  \apjl, 904, L18, \dodoi{10.3847/2041-8213/abc704}

\bibitem[{{Sakai} {et~al.}(2014){Sakai}, {Oya}, {Sakai}, {Watanabe}, {Hirota},
  {Ceccarelli}, {Kahane}, {Lopez-Sepulcre}, {Lefloch}, {Vastel}, {Bottinelli},
  {Caux}, {Coutens}, {Aikawa}, {Takakuwa}, {Ohashi}, {Yen}, \&
  {Yamamoto}}]{Sakai14b}
{Sakai}, N., {Oya}, Y., {Sakai}, T., {et~al.} 2014, \apjl, 791, L38,
  \dodoi{10.1088/2041-8205/791/2/L38}

\bibitem[{{Schwarz} {et~al.}(2016){Schwarz}, {Bergin}, {Cleeves}, {Blake},
  {Zhang}, {{\"O}berg}, {van Dishoeck}, \& {Qi}}]{Schwarz16}
{Schwarz}, K.~R., {Bergin}, E.~A., {Cleeves}, L.~I., {et~al.} 2016, \apj, 823,
  91, \dodoi{10.3847/0004-637X/823/2/91}

\bibitem[{{Schwarz} {et~al.}(2018){Schwarz}, {Bergin}, {Cleeves}, {Zhang},
  {{\"O}berg}, {Blake}, \& {Anderson}}]{Schwarz18}
---. 2018, \apj, 856, 85, \dodoi{10.3847/1538-4357/aaae08}

\bibitem[{{Teague} {et~al.}(2018){Teague}, {Bae}, {Bergin}, {Birnstiel}, \&
  {Foreman-Mackey}}]{Teague18}
{Teague}, R., {Bae}, J., {Bergin}, E.~A., {Birnstiel}, T., \& {Foreman-Mackey},
  D. 2018, \apj, 860, L12, \dodoi{10.3847/2041-8213/aac6d7}

\bibitem[{{Tobin} {et~al.}(2018){Tobin}, {Looney}, {Li}, {Sadavoy}, {Dunham},
  {Segura-Cox}, {Kratter}, {Chandler}, {Melis}, {Harris}, \& {Perez}}]{Tobin18}
{Tobin}, J.~J., {Looney}, L.~W., {Li}, Z.-Y., {et~al.} 2018, \apj, 867, 43,
  \dodoi{10.3847/1538-4357/aae1f7}

\bibitem[{{Toomre}(1964)}]{Toomre64}
{Toomre}, A. 1964, \apj, 139, 1217, \dodoi{10.1086/147861}

\bibitem[{{Trapman} {et~al.}(2019){Trapman}, {Facchini}, {Hogerheijde}, {van
  Dishoeck}, \& {Bruderer}}]{Trapman19}
{Trapman}, L., {Facchini}, S., {Hogerheijde}, M.~R., {van Dishoeck}, E.~F., \&
  {Bruderer}, S. 2019, \aap, 629, A79, \dodoi{10.1051/0004-6361/201834723}

\bibitem[{{Trapman} {et~al.}(2017){Trapman}, {Miotello}, {Kama}, {van
  Dishoeck}, \& {Bruderer}}]{Trapman17}
{Trapman}, L., {Miotello}, A., {Kama}, M., {van Dishoeck}, E.~F., \&
  {Bruderer}, S. 2017, \aap, 605, A69, \dodoi{10.1051/0004-6361/201630308}

\bibitem[{{van der Marel} {et~al.}(2018){van der Marel}, {Williams}, {Ansdell},
  {Manara}, {Miotello}, {Tazzari}, {Testi}, {Hogerheijde}, {Bruderer}, {van
  Terwisga}, \& {van Dishoeck}}]{vanderMarel18}
{van der Marel}, N., {Williams}, J.~P., {Ansdell}, M., {et~al.} 2018, \apj,
  854, 177, \dodoi{10.3847/1538-4357/aaaa6b}

\bibitem[{{van der Walt} {et~al.}(2011){van der Walt}, {Colbert}, \&
  {Varoquaux}}]{numpy}
{van der Walt}, S., {Colbert}, S.~C., \& {Varoquaux}, G. 2011, Computing in
  Science and Engineering, 13, 22, \dodoi{10.1109/MCSE.2011.37}

\bibitem[{{Weaver} {et~al.}(2018){Weaver}, {Isella}, \& {Boehler}}]{Weaver18}
{Weaver}, E., {Isella}, A., \& {Boehler}, Y. 2018, \apj, 853, 113,
  \dodoi{10.3847/1538-4357/aaa481}

\bibitem[{{Williams} \& {Best}(2014)}]{Williams14}
{Williams}, J.~P., \& {Best}, W.~M.~J. 2014, \apj, 788, 59,
  \dodoi{10.1088/0004-637X/788/1/59}

\bibitem[{{Wilson}(1999)}]{Wilson99}
{Wilson}, T.~L. 1999, Reports on Progress in Physics, 62, 143,
  \dodoi{10.1088/0034-4885/62/2/002}

\bibitem[{{Woitke} {et~al.}(2016){Woitke}, {Min}, {Pinte}, {Thi}, {Kamp},
  {Rab}, {Anthonioz}, {Antonellini}, {Baldovin-Saavedra}, {Carmona}, {Dominik},
  {Dionatos}, {Greaves}, {G{\"u}del}, {Ilee}, {Liebhart}, {M{\'e}nard},
  {Rigon}, {Waters}, {Aresu}, {Meijerink}, \& {Spaans}}]{Woitke16}
{Woitke}, P., {Min}, M., {Pinte}, C., {et~al.} 2016, \aap, 586, A103,
  \dodoi{10.1051/0004-6361/201526538}

\bibitem[{{Woitke} {et~al.}(2019){Woitke}, {Kamp}, {Antonellini}, {Anthonioz},
  {Baldovin-Saveedra}, {Carmona}, {Dionatos}, {Dominik}, {Greaves},
  {G{\"u}del}, {Ilee}, {Liebhardt}, {Menard}, {Min}, {Pinte}, {Rab}, {Rigon},
  {Thi}, {Thureau}, \& {Waters}}]{Woitke19}
{Woitke}, P., {Kamp}, I., {Antonellini}, S., {et~al.} 2019, \pasp, 131, 064301,
  \dodoi{10.1088/1538-3873/aaf4e5}

\bibitem[{{Woodall} {et~al.}(2007){Woodall}, {Ag{\'u}ndez}, {Markwick-Kemper},
  \& {Millar}}]{Woodall07}
{Woodall}, J., {Ag{\'u}ndez}, M., {Markwick-Kemper}, A.~J., \& {Millar}, T.~J.
  2007, \aap, 466, 1197, \dodoi{10.1051/0004-6361:20064981}

\bibitem[{{Xu} {et~al.}(2017){Xu}, {Bai}, \& {{\"O}berg}}]{Xu17}
{Xu}, R., {Bai}, X.-N., \& {{\"O}berg}, K. 2017, \apj, 835, 162,
  \dodoi{10.3847/1538-4357/835/2/162}

\bibitem[{{Yen} {et~al.}(2019){Yen}, {Gu}, {Hirano}, {Koch}, {Lee}, {Liu}, \&
  {Takakuwa}}]{Yen19}
{Yen}, H.-W., {Gu}, P.-G., {Hirano}, N., {et~al.} 2019, \apj, 880, 69,
  \dodoi{10.3847/1538-4357/ab29f8}

\bibitem[{{Yu} {et~al.}(2016){Yu}, {Willacy}, {Dodson-Robinson}, {Turner}, \&
  {Evans}}]{Yu16}
{Yu}, M., {Willacy}, K., {Dodson-Robinson}, S.~E., {Turner}, N.~J., \& {Evans},
  II, N.~J. 2016, \apj, 822, 53, \dodoi{10.3847/0004-637X/822/1/53}

\bibitem[{{Zhang} \& {MAPS Team}(2021)}]{ZhangMAPS}
{Zhang}, \& {MAPS Team}. 2021, \apj, 0, 0, \dodoi{0}

\bibitem[{{Zhang} {et~al.}(2017){Zhang}, {Bergin}, {Blake}, {Cleeves}, \&
  {Schwarz}}]{Zhang17}
{Zhang}, K., {Bergin}, E.~A., {Blake}, G.~A., {Cleeves}, L.~I., \& {Schwarz},
  K.~R. 2017, Nature Astronomy, 1, 0130, \dodoi{10.1038/s41550-017-0130}

\end{thebibliography}
\bibliographystyle{aasjournal}

\end{document}